\documentclass[12pt,a4paper]{article}
\usepackage[authoryear]{natbib}
\usepackage{times}
\usepackage{graphics}
\usepackage{ifpdf}
\usepackage{graphicx}
\usepackage{pgf}
\usepackage{epsfig}
\usepackage{amssymb}
\usepackage{amsmath,latexsym,xspace}
\usepackage{array}
\usepackage{setspace} 
\usepackage{comment}
\usepackage{authblk}
\usepackage{mathtools}
\usepackage{float}
\usepackage{booktabs}
\usepackage[scale={0.76,0.87},centering,includeheadfoot]{geometry}

\def\mm#1{\ensuremath{\boldsymbol{#1}}}

\setkeys{Gin}{width=0.45\textwidth}
\begin{document}
\begin{titlepage}
\title{Bayesian identification of early warning signals for long-range dependent climatic time series}

\author[1]{Sigrunn Holbek S\o rbye$^1$, Eirk Myrvoll-Nilsen}
\author[2]{H\aa vard Rue}
\affil[1]{Department of Mathematics and Statistics, UiT The Arctic University of Norway, 9037 Troms{\o}, Norway}
\affil[2]{CEMSE Division, King Abdullah University of Science and Tecnhology, Saudi Arabia}
\end{titlepage}
\maketitle
\begin{abstract}

Detecting early warning signals in climatic time series is essential for  anticipating critical transitions and tipping points. Common statistical indicators include increased variance and lag-one autocorrelation prior to bifurcation points. However, these indicators are sensitive to observational noise, long-term mean trends, and long-memory dependence, all of which are prevalent in climatic time series. Such effects can easily obscure genuine signals or generate spurious detections. To address these challenges, we employ a flexible Bayesian framework for modelling time-varying autocorrelation in long-range dependent time series, also accounting for time-varying variance. The approach uses a mixture of two fractional Gaussian noise processes with a time-dependent weight function to represent fractional Gaussian noise with a time-varying Hurst exponent. Inference is performed via integrated nested Laplace approximation, enabling joint estimation of mean trends and handling of irregularly sampled observations. The strengths and limitations of detecting changes in the autocorrelation is investigated in extensive simulations. Applied to real climatic data sets, we find evidence of  early warning signals in a reconstructed Atlantic multidecadal variability index, while dismissing such signals for the paleoclimate records spanning the Dansgaard–Oeschger events. 
\end{abstract}

{\bf Keywords:} Bayesian inference, Dansgaard-Oeschger events, fractional Gaussian noise, INLA, time-varying autocorrelation, tipping points

\section{Introduction}
Global climate change is expected to increase the risk of abrupt and potentially irreversible transitions in the Earth system. Of particular concern are tipping points associated with the Greenland and West Antarctic ice sheets, permafrost thawing, Amazon rainforest die-back, coral reef degradation, and the Atlantic Meridional Overturning Circulation (AMOC) \citep{lenton:25, mckay:22}.

Numerous early warning indicators have been proposed to assess whether a dynamical system is approaching a tipping point, see \citet{dakos:24} for an overview. Many such indicators are motivated by the theory of critical slowing down (CSD), which characterises bifurcation-induced tipping points. Specifically, as control parameters approach critical threshold values, the system would loose resilience and recover more slowly from small perturbations \citep{wissel:84}. Simple statistical indicators of CSD include increasing variance and lag-one autocorrelation, patterns which have been reported prior to abrupt climate shifts in paleoclimate records \citep{dakos:08, scheffer:09}. These indicators are commonly estimated using sliding windows and assessed using Kendall's rank correlation coefficient \citep{kendall:38}.  However, their performance is sensitive to the choice of window size and temporal resolution \citep{chen:22}. Moreover, changes in second-moment properties are easily confounded by observational noise and preprocessing steps such as detrending, filtering and interpolation. Also, these indicators might be less effective for time series with long-range dependence, a characteristic feature of many climatic time series \citep{franzke2012nonlinear, koscielny1998indication, mei:24, yuan:15}. 

To address these challenges and extend on current methodology, this paper focuses on identifying changes in the autocorrelation of long-range dependent time series, also accounting for time-varying marginal variance. Specifically, we propose a novel approach for modelling time-varying autocorrelation for long-range dependent time series, having properties in coherence with fractional Gaussian noise (fGn) \citep{hurst:51}. This is achieved by mixing two independent and scaled fGn processes using a time-dependent weight function. By making use of the Kullback-Leibler divergence (KLD), the weight function is mapped to a time-varying Hurst exponent that encodes the memory properties of the series. This formulation provides a conceptually simple and computationally efficient alternative to existing approaches modelling a time-varying Hurst exponent directly \citep{ryvkina:15}. 

While the global Hurst exponent characterises dependence across all time scales, it does not capture local instability of a system approaching a bifurcation point. However, an increase in time-varying (local) Hurst exponents is consistent with increased correlation time at higher frequencies, indicating loss of stability \citep{rypdal:16}. This has motivated the use of local  Hurst exponents as early warning indicators, for example via wavelet-based methods and sliding window estimates \citep{boers:18, hummel:25, rypdal:16}. \citet{mei:24} showed that local Hurst exponents may be more robust to spurious external signals than indicators based on variance and lag-one autocorrelation in folded bifurcation models. 

Our approach extends on previous analyses for long-range dependent time series, by embedding the new model formulation in a Bayesian hierarchical framework. The proposed model is incorporated within the class of latent Gaussian models, fitted using the methodology of integrated nested Laplace approximations (INLA) \citep{rueal:09}. This allows for joint inference of time-varying autocorrelation and variance, and additional latent components such as mean trends. In contrast to two-step approaches that rely on prior detrending, this ensures coherent uncertainty propagation, also offering flexibility to handle irregularly sampled time series. The given approach also extends on the studies of \citet{myrvoll1:25} and \citet{myrvoll2:25} who used a similar Bayesian framework to identify increased lag-one  autocorrelation of short-term memory processes, but not accounting for separate time-varying marginal variance.

The remainder of the paper is structured as follows. Section~\ref{sec:2} provides details on the proposed model, including the mapping between the weight function and the time-varying Hurst exponent, as well as implementation within a Bayesian framework. Section~\ref{sec:3} investigates inferential properties and the classification performance through extensive simulations.  Section~\ref{sec:4} applies the new model to investigate early warning signals of a millennial Atlantic multidecadal variability reconstruction provided by  \citet{michel:22}. The section also investigates the time series spanning the abrupt paleoclimate shifts referred to as the Dansgaard–Oeschger events \citep{dansgaard:84, dansgaard:93}, dismissing early warning signals in terms of increased autocorrelation for these series. Discussion and concluding remarks are given in Section~\ref{sec:5}. 

\section{Methodology} \label{sec:2}
\subsection{Background on fGn}
An fGn is defined as the discrete-time stationary increment process of fractional Brownian motion, originally developed in the hydrological sciences \citep{hurst:51, hurst:57, mandelbrot:68}. Since its introduction, this model has been used in a wide range of
applications, see e.g.  \citet{graves:17} for a comprehensive review.  
 A key advantage of fGn is its parsimonious parametrisation through the Hurst exponent $H$, which characterises the dependency structure of the process.  
The autocorrelation function (ACF) is defined as  
$$\rho_H(k) =  \frac{1}{2}\left(|k+1|^{2H} -2|k|^{2H}+|k-1|^{2H}\right), \quad k=0,1,\ldots ,n-1, $$
where $n$ denotes the length of the series. Asymptotically, the autocorrelation function exhibits a power law decay 
\begin{equation}
\lim_{k\rightarrow \infty} \rho_H(k) \sim H(2H-1)k^{2(H-1)}, \label{eq:power-law}
\end{equation}
implying long-range dependence when $H\in (0.5, 1)$. The fGn has anti-persistent behaviour when $H\in (0, 0.5)$, while reducing to white noise when $H=0.5$. The covariance matrix $\Sigma_H$ of fGn is Toeplitz, allowing likelihood-based inference for a series of length $n$ to be computed with a  cost of ${\mathcal O}(n^2)$ \citep{mcleod:07}. However, the Toeplitz structure is typically disrupted when the fGn is observed indirectly or with inhomogeneous noise, in which case the computational cost increases to cubic order \citep{beran:94, sorbye:19}.  

\subsection{Novel  model incorporating time-varying autocorrelation}
Let $\mm{x}_{H_1}$ and $\mm{x}_{H_2}$ denote independent and scaled fGn processes with Hurst exponents $H_1$ and $H_2$, respectively. To incorporate time-varying autocorrelation, we define  a mixed model component specified by 
\begin{equation}
\epsilon(t_i) =\tau^{-1/2} \left(\sqrt{1-w(t_i)}x_{H_1}(t_i) + \sqrt{w(t_i)}x_{H_2}(t_i)\right), \quad \quad i=1,\ldots , n.\label{eq:model}
\end{equation}
The model formulation includes a deterministic linear weight function,
\begin{equation}
w(t_i) = (t_i-t_1)/(t_n-t_1) \label{eq:weight}
\end{equation}
where the locations $t_1,\ldots , t_n$, for now,  are assumed to be equally spaced. 
By definition, the proposed model component has zero mean and constant variance $\sigma^2 = \tau^{-1}$, however this will be relaxed when needed, as explained  in Section \ref{sec:4.2.2}. Importantly, the autocorrelation for the mixed model component varies in time according to the weight function. Specifically, the elements of the covariance matrix $\mm{\Sigma}_{\epsilon}$ of  $\mm{\epsilon}  =(\epsilon(t_1),\ldots , \epsilon(t_n))^{\top}$ 
are given by 
\begin{eqnarray*}
\sigma^2_{\epsilon,ij}  
 & = &  \tau^{-1}\left[(1-w(t_i))\rho_{H_1}(|i-j|) + w(t_i)\rho_{H_2}(|i-j|)\right]
\end{eqnarray*}
for all $i,j  =  1,\ldots , n$.   

\begin{figure}
\includegraphics[width=0.325\textwidth]{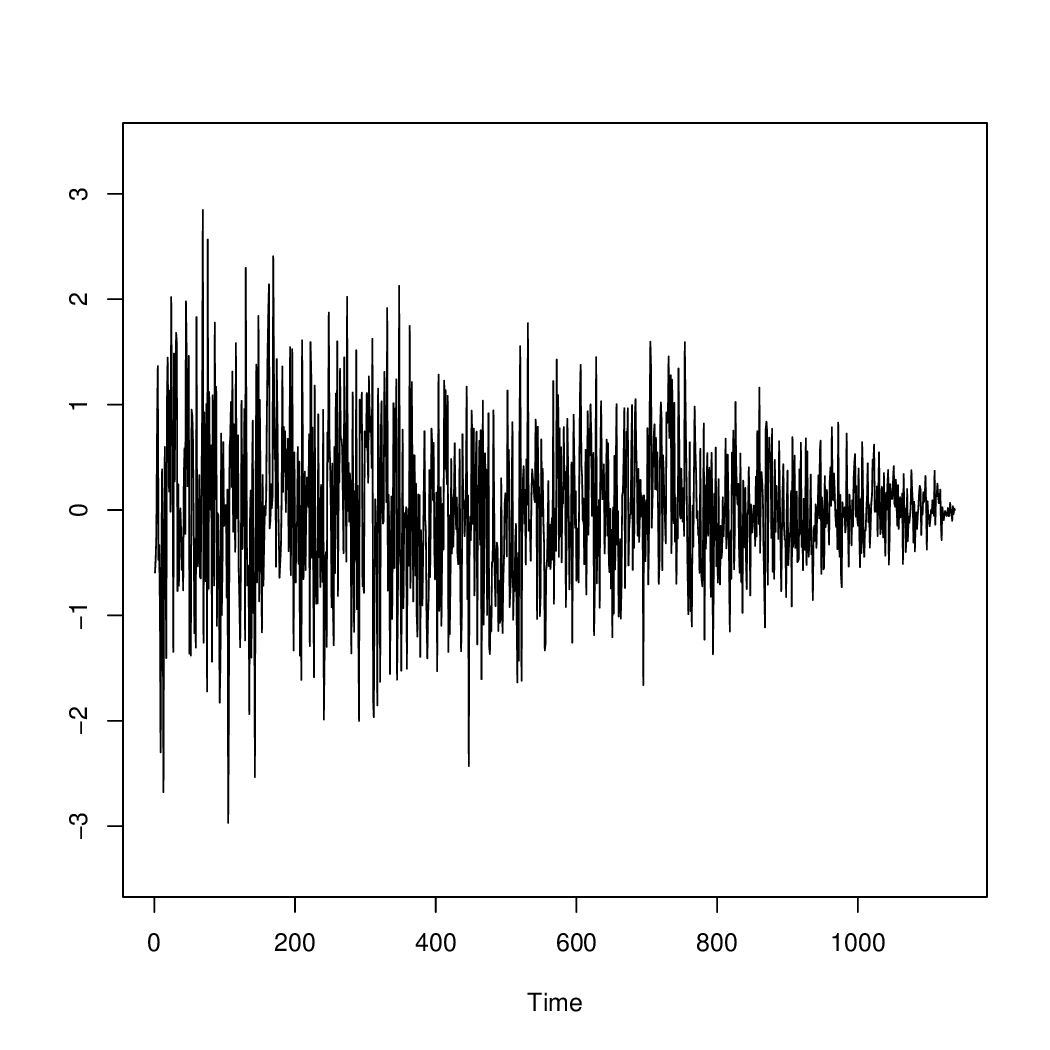}
\includegraphics[width=0.325\textwidth]{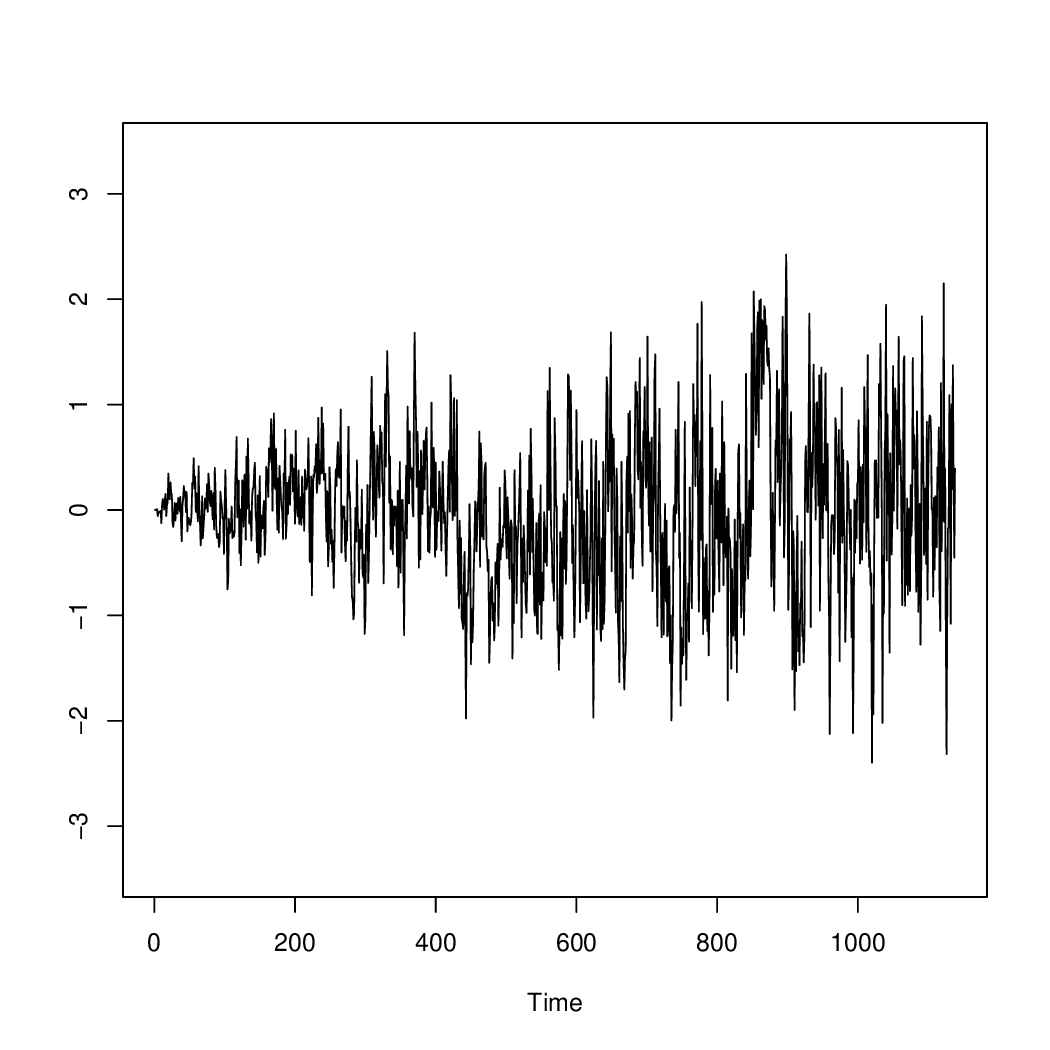}
\includegraphics[width=0.325\textwidth]{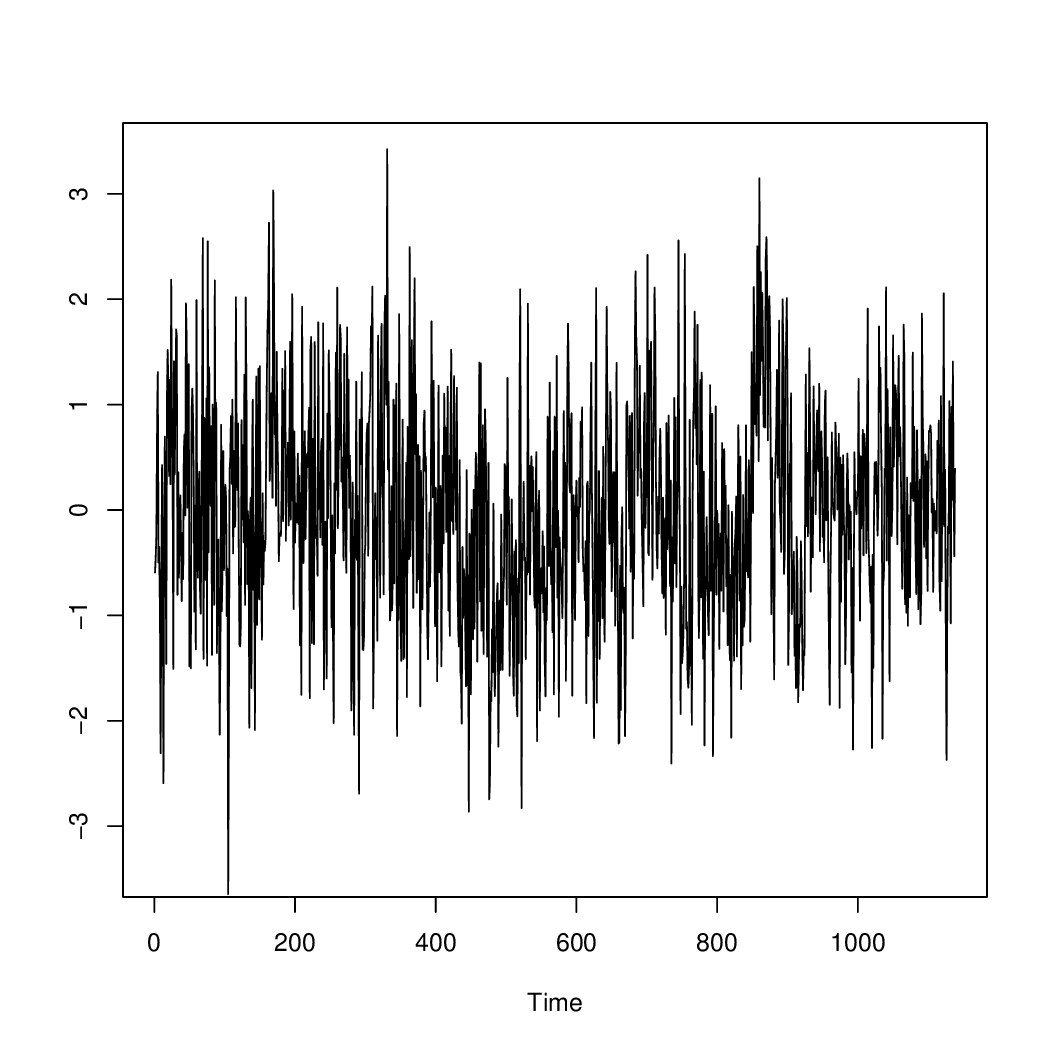}
\caption{Simulated realization from \eqref{eq:model} illustrating the weighted sum of  two fGns having Hurst exponents $H_1 = 0.6 $ (left)  and $H_2 = 0.8$ (middle), giving a process with time-varying autocorrelation (right).}
\label{fig:H0608}
\end{figure}

A simulated example illustrating the  construction of the proposed model  component is given in Figure~\ref{fig:H0608}. We first generate two independent and scaled fGn processes having Hurst exponents $H_1 = 0.6$ and $H_2=0.8$, respectively. These are mixed using the linear weight function in \eqref{eq:weight}, providing a realization from \eqref{eq:model} with increasing autocorrelation. 

\subsection{Mapping between the weight function and time-varying Hurst exponent}
The mixed model component in \eqref{eq:model} exhibits long-range dependence, as its autocorrelation function follows a power law decay with Hurst exponent $H_2$ (see Appendix \ref{app:asymptotic}). This motivates approximating the new model with a single fGn process with local Hurst exponents $H(t_i)$ for each time point $t_i$.  In particular, this approximation allows us to construct a mapping between the weights $w(t_i)$ and the corresponding Hurst exponents $H(t_i)$, effectively mimicking an fGn process with time-varying local dependence. 

To determine an optimal value of $H(t_i)$ for each time point,  we minimize the KLD between the  density $f_{\mm{\epsilon}}$ of the mixed model component and  the approximating density $f_H$ of fGn. This divergence is given by 
\begin{eqnarray*}
\mbox{KLD}(f_{\mm{\epsilon}}\parallel f_H) &=& \int f_{\mm{\epsilon}}(\mm{u}) \log \left(\frac{f_{\mm{\epsilon}}(\mm{u})}{ f_H(\mm{u})}\right)d\mm{u}=  -\frac{1}{2} \sum_{j=1}^n \left(\log(\frac{\lambda_j}{q_j}) - \frac{\lambda_j}{q_j}  +1\right), 
\end{eqnarray*}
where $\lambda_j$ and $q_j$ denote the  eigenvalues of the covariance matrices $\mm{\Sigma}_{\epsilon}$ and $\mm{\Sigma}_H$, respectively.  For a fixed weight,  both $\mm{\Sigma}_{\epsilon}$ and  $\mm{\Sigma}_H$ are Toeplitz, allowing the eigenvalues to be computed efficiently using  the  first row of these covariance matrices. This is achieved by making the corresponding vectors circular and calculating the eigenvalues through the discrete fast Fourier transform \citep{rue:02}. 

\begin{figure}[h]
\centering
\includegraphics[width = 0.4\textwidth]{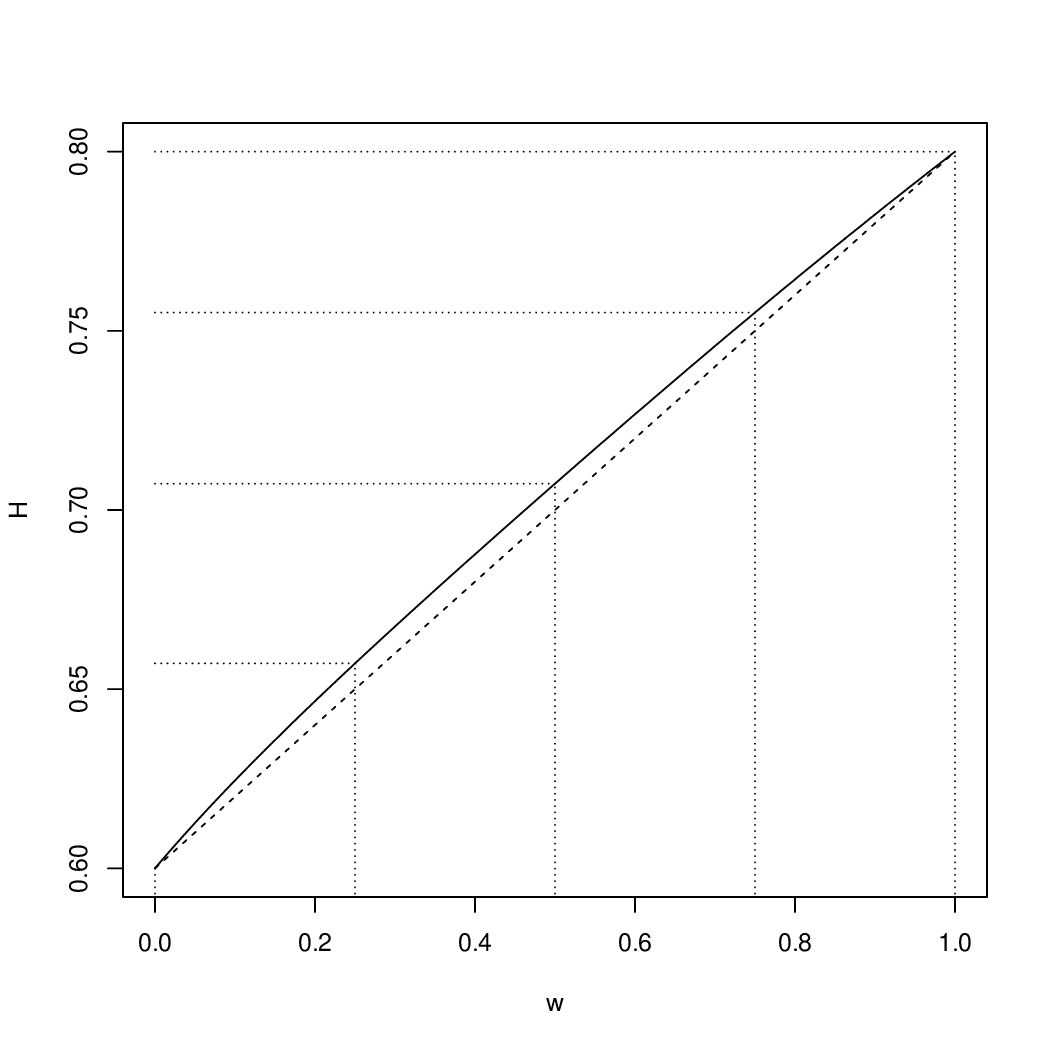}
\caption{Mapping (solid line) between the Hurst exponent and the weight function found by minimizing the KLD between an fGn and the proposed model component in \eqref{eq:model} with $H_1 = 0.6$ and $H_2 = 0.8$.} 
\label{fig:mapping}
\end{figure}

The  resulting mapping between the  linear weight function and the parameter $H$,  obtained by minimizing KLD, is shown in Figure~\ref{fig:mapping} where $H_1 = 0.6 $ and $H_2 = 0.8$ in~\eqref{eq:model}. The mapping is very close to linear and will approach exact linearity as $|H_1-H_2|$ decreases. 
This suggests that the weight function can readily be adapted to represent nonlinear variation in the Hurst exponent, if needed.

\subsection{Implementation within a Bayesian framework}
In fitting the proposed model to a time series, the primary objective is to estimate the Hurst exponents  $H_1$ and $H_2$ in order to assess whether the autocorrelation is increasing. We  employ a Bayesian framework and  perform inference using INLA, available through the \texttt{R}-package \texttt{R-INLA} \citep{rueal:09, rue:17}. The new model component is formulated within the general class of latent Gaussian models, which has  a three-stage hierarchical structure. These stages include hyperparameters $\mm{\theta}$, a latent random field  $\mm{\eta}$, and the observations $\mm{y}$. 

The proposed model has three hyperparameters,  $\mm{\theta} = (\tau , H_1, H_2)^{\top}$. Both the precision parameter and the two Hurst exponents are assigned penalised complexity priors as described in \citet{simpson:17} and \citet{sorbye:18}, respectively.  In short, these priors are  derived based on specific principles which induce parsimonious models,  effectively controlling shrinkage towards simpler base models. The rate of shrinkage is tuned in an  intuitive way by the user, reducing the risk of overfitting the model to the data while maintaining the ability to capture complex dynamics.

Conditioned on  $\mm{\theta}$,  the  latent field $\mm{\eta}$ is assigned a multivariate Gaussian prior distribution
$$\mm{\eta}\mid \mm{\theta}\sim \pi_G(0, \mm{Q}^{-1}(\mm{\theta})),$$
where $\mm{Q}(.)^{-1}$ denotes the precision (inverse covariance) matrix. If the model structure only includes the new  model component,   the latent random field is given by $\mm{\eta} =\mm{\epsilon}$.  However, we can easily add other latent model components to $\mm{\eta}$, for example including a linear or non-linear mean trend. Finally, the observed time series  $\mm{y}= (y_1,\ldots , y_n)^\top$ is assumed conditionally independent given $\mm{\eta}$ and  $\mm{\theta}$, leading to the  likelihood
$$L(\mm{\eta},\mm{\theta}\mid \mm{y}) = \prod_{i=1}^n \pi(y_i\mid \eta_i, \mm{\theta})$$
which here is assumed Gaussian. 
In summary, the joint posterior distribution for all components of the latent field and all hyperparameters is defined by
$$\pi(\mm{\eta}, \mm{\theta}\mid \mm{y})\propto \prod_{i=1}^n \pi(y_i\mid \eta_i, \mm{\theta})\pi(\mm{\eta}\mid \mm{\theta})\pi(\mm{\theta}).$$ 
The INLA-methodology provides accurate approximations of the marginal posterior distributions of all components of the latent field and all hyperparameters.

Ordinarily, the latent field $\mm{\eta}$ is assumed to be a Gaussian Markov random field (GMRF) having a sparse precision matrix. However,  due to the long-range dependency structure of  fGn when $H>0.5$, this process is not a GMRF. To overcome this limitation, we make use of the built-in representation of fGn in \texttt{R-INLA}, defined by a weighted sum of $m = 3$ or $m=4$ independent first-order autoregressive processes (AR(1)),
\begin{equation}
\tilde{\mm{x}}_{H} = \sum_{j=1}^m \sqrt{v_{j}}\mm{z}_j. \label{eq:ar-approx}
\end{equation}
 The set $\{\mm{z}_j\}_{j=1}^m$ denote independent and scaled AR(1) processes. The lag-one autocorrelation coefficients of these processes and the weights $\{v_{j}\}_{j=1}^m$ are predefined to minimize the absolute difference between the ACF of the AR-approximation and the exact fGn,  for each $H\in (0.50, 0.99)$. Due to Markov properties, this representation substantially reduces the potential cost of fGn from cubic to linear, as well as being very accurate \citep{sorbye:19}. The Markov representation of the proposed model \eqref{eq:model} is  then defined by  
\begin{equation}
\tilde{\epsilon}(t_i)= \tau^{-1/2}\left(\sqrt{1-w(t_i)} \tilde x_{H_1}(t_i) + \sqrt{w(t_i)} \tilde x_{H_2}(t_i)\right)
\label{eq:approx-model}
\end{equation}
This model is implemented as a new latent model component using the \texttt{rgeneric} framework which accompanies the \texttt{R-INLA} package, see Appendix~\ref{app:implementation} for further details. 

\section{Simulation results} \label{sec:3}
This section investigates the properties of the proposed model defined in~\eqref{eq:approx-model} using extensive simulations.  We first assess  the  accuracy of parameter estimates for fixed combinations of the Hurst exponents $H_1$ and $H_2$. Second, we examine the model's classification performance by  generating realizations from the new model component with randomly chosen Hurst exponents in $(0.5,1)$. The main objective of the latter analysis is to investigate the models ability to detect early warning signals in terms of increased autocorrelation versus the need to control for false positive findings.

\begin{table}[h]
\begin{center}
\begin{tabular}{rrrrrrrr}
\toprule
Length & $H_1 $ & $H_2$ & $\hat H_1$ &  $\hat H_2$ &  $\widehat{\mbox{RMSE}}$$_1$ &  $\widehat{\mbox{RMSE}}_2$ & $\hat p$ \\
\midrule
$n = 200$ &0.6 & 0.6  & 0.617 & 0.616 & 0.049 & 0.047 & 0.506 \\
&0.7 & 0.7  & 0.685 & 0.686 & 0.061 & 0.062 & 0.497  \\
&0.8 & 0.8  & 0.768 & 0.768 & 0.068 & 0.067 & 0.510 \\
&0.9 & 0.9  & 0.836 & 0.837 & 0.081 & 0.080 & 0.494  \\  
 &0.6 & 0.7 & 0.629 & 0.675 & 0.057 & 0.063  & 0.730\\
&0.7 & 0.8 & 0.700 & 0.756 & 0.059 & 0.075  & 0.742  \\
&0.8 & 0.9 & 0.774 & 0.824 &  0.061 & 0.092 & 0.761 \\ 
&0.6 & 0.8 & 0.635 & 0.737 & 0.061 & 0.087  & 0.887  \\
&0.7 & 0.9 & 0.714 & 0.821 & 0.060 & 0.097  & 0.921 \\ 
&0.6 & 0.9 & 0.646 & 0.805 & 0.067 & 0.111 &  0.978\\\hline
$n = 500$ & 0.6 & 0.6 & 0.605 & 0.607 & 0.038 & 0.039 &  0.511\\
&0.7 & 0.7 & 0.692 & 0.690 & 0.045 & 0.047 & 0.494\\
&0.8 & 0.8 & 0.779 & 0.779 & 0.045 & 0.046 & 0.491 \\
&0.9 & 0.9 & 0.863 & 0.863 & 0.049 & 0.044 & 0.507 \\
&0.6 & 0.7 & 0.612 & 0.684 & 0.043& 0.047  & 0.849 \\
&0.7 & 0.8 & 0.698 & 0.773 & 0.044 & 0.049 & 0.869 \\
&0.8 & 0.9 & 0.791 & 0.853 & 0.039 & 0.058 & 0.890  \\
&0.6 & 0.8 & 0.617 & 0.767 & 0.043 & 0.054 & 0.981 \\
&0.7 & 0.9 & 0.708 & 0.846 & 0.042 & 0.065 & 0.987 \\
&0.6 & 0.9 & 0.626 & 0.840 & 0.048 & 0.071 & 1.000 \\\hline 
$n = 1000$ & 0.6 & 0.6 & 0.601 & 0.603 & 0.032 & 0.032 & 0.527 \\
& 0.7 & 0.7 & 0.698 & 0.695 & 0.034   & 0.036  & 0.486  \\
& 0.8 & 0.8 & 0.792 & 0.793 & 0.030 & 0.029 &  0.518 \\
& 0.9 & 0.9 & 0.887 & 0.888& 0.028 & 0.028 & 0.489   \\
& 0.6 & 0.7 & 0.602 & 0.679  & 0.032 & 0.038 & 0.923  \\
& 0.7 & 0.8 & 0.705 & 0.784 & 0.033 & 0.037   & 0.951  \\ 
& 0.8 & 0.9 & 0.808 & 0.872 & 0.029  & 0.036 & 0.972    \\
& 0.6 & 0.8 & 0.613 & 0.783 & 0.036 & 0.038 &  1.000 \\ 
& 0.7 & 0.9 & 0.711 & 0.859 &  0.033&  0.049 & 1.000 \\
&0.6 & 0.9 & 0.625 & 0.859 & 0.042 & 0.049 & 1.000\\\hline
\end{tabular}
\caption{Posterior mean estimates of the Hurst exponents, frequentist estimates of RMSE and the proportion $\hat p =  \#\{\hat H_2 > \hat H_1\}/N$. The results are averaged over $N= 1000$ simulations for each fixed combination of the two Hurst exponents and for each time series lengths $n=200, 500, 1000$.  \label{tab:est-all}}
\end{center}
\end{table}

\clearpage
\subsection{Inferential properties}
Realizations from the proposed mixed model are generated using \eqref{eq:model} with fixed combinations of the Hurst exponents $H_1$ and $H_2$. For each combination, we simulate $N=1000$ time series, and we repeat the experiment for series lengths $n=200, 500$ and $1000$. We then fit the model defined in \eqref{eq:approx-model} to obtain posterior mean estimates of the Hurst exponent. In addition, we report the frequentist root mean squared error (RMSE) for estimating the Hurst exponents, along with the proportion of simulations in which $\hat H_2>\hat H_1$ (Table~\ref{tab:est-all}).

We first assess performance for stationary time series, considering the four combinations in which $H_1 = H_2$. As expected for these cases, the proportion of simulations in which $\hat H_2>\hat H_1$ is approximately 0.5. The accuracy of the parameter estimates decreases as the true Hurst exponents increase, the estimates being biased downwards. The bias is particularly  evident for the shortest series with $n=200$, giving RMSE values in the range $(0.047, 0.081)$. This indicates that series of this length do not contain sufficient information to produce more accurate estimates. The downward bias is noticeably reduced for time series of lengths $n=500$, where the RMSE falls within the range $(0.038, 0.049)$. For time series of length $n=1000$, the parameter estimates are very accurate, having RMSE values in the range $(0.028, 0.032)$.  

Second, we evaluate the six combinations of Hurst exponents in which $H_2>H_1$. These results show that the upper Hurst exponent $H_2$ is systematically underestimated, yielding a larger RMSE than the estimates of $H_1$ in all cases. This is not surprising, as only the last observation of the series is constructed to have a Hurst exponent equal to $H_2$.  The estimation error increases with the magnitude of $H_2$. However, this is compensated by the fact that the probability of correctly detecting an increase in the Hurst exponent also increases with the true value. This makes the downward bias less problematic using the model as an early warning indicator of increased autocorrelation. We notice that the probability of detecting an increase in the Hurst exponent exceeds 0.95 for only one combination when $n = 200$. This increases to three  combinations when $n=500$. When $n=1000$, the probability of detecting an increase in the Hurst exponent exceeds 0.95 for five of the combinations (Table \ref{tab:est-all}). 

\subsection{Performance in terms of classification}
Prediction of whether a time series exhibits early warning signal in terms of increased autocorrelation  can be viewed as a binary classification problem, in which the probability of positive detection must be balanced against the risk of false alarms. To study this trade-off, we generate 2000 time series from \eqref{eq:model}, with $H_1$ and $H_2$ drawn independently and randomly from the interval $(0.50, 0.99)$. Fitting the proposed model \eqref{eq:approx-model} to these series, increased autocorrelation is predicted based on  the posterior probability  $\hat p = P(H_2-H_1>0\mid \mm{y})$. This probability is computed efficiently  by  Monte Carlo simulation from  the joint posterior distribution of the hyperparameters provided by  \texttt{R-INLA}. 

\begin{figure}[ht]
\begin{center}
\includegraphics[width=0.32\textwidth]{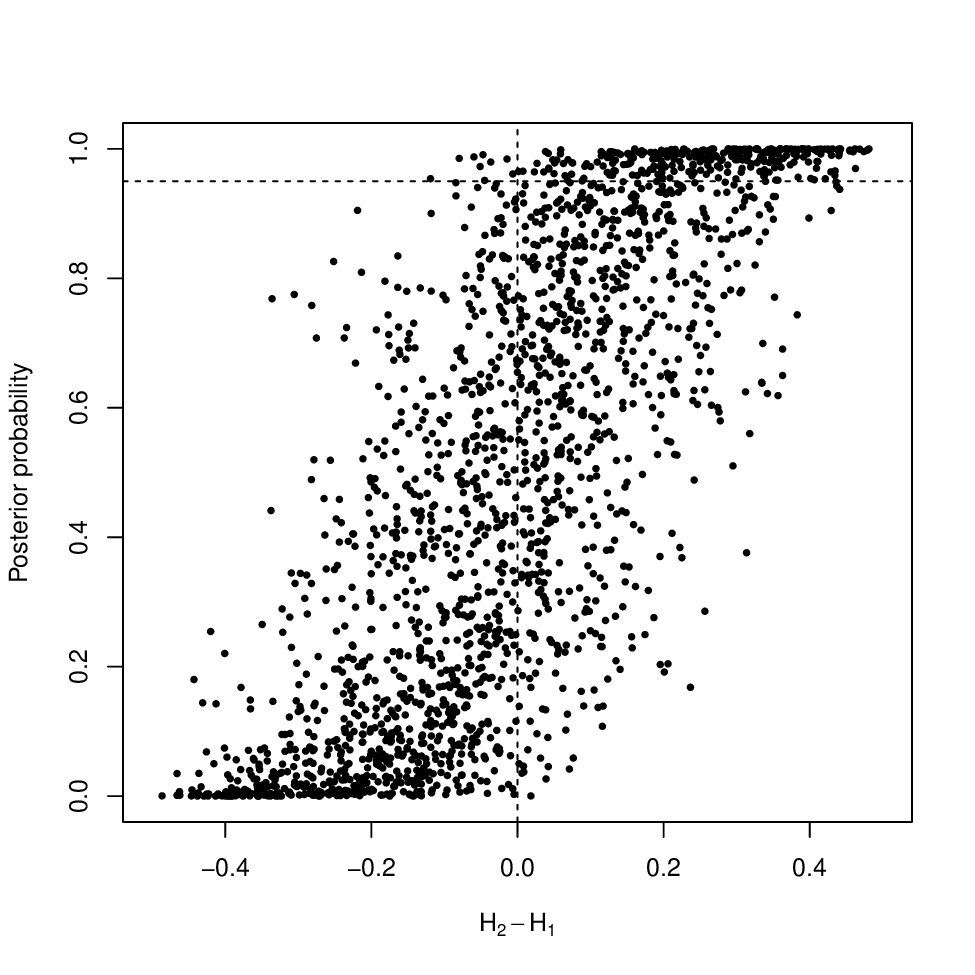} 
\includegraphics[width=0.32\textwidth]{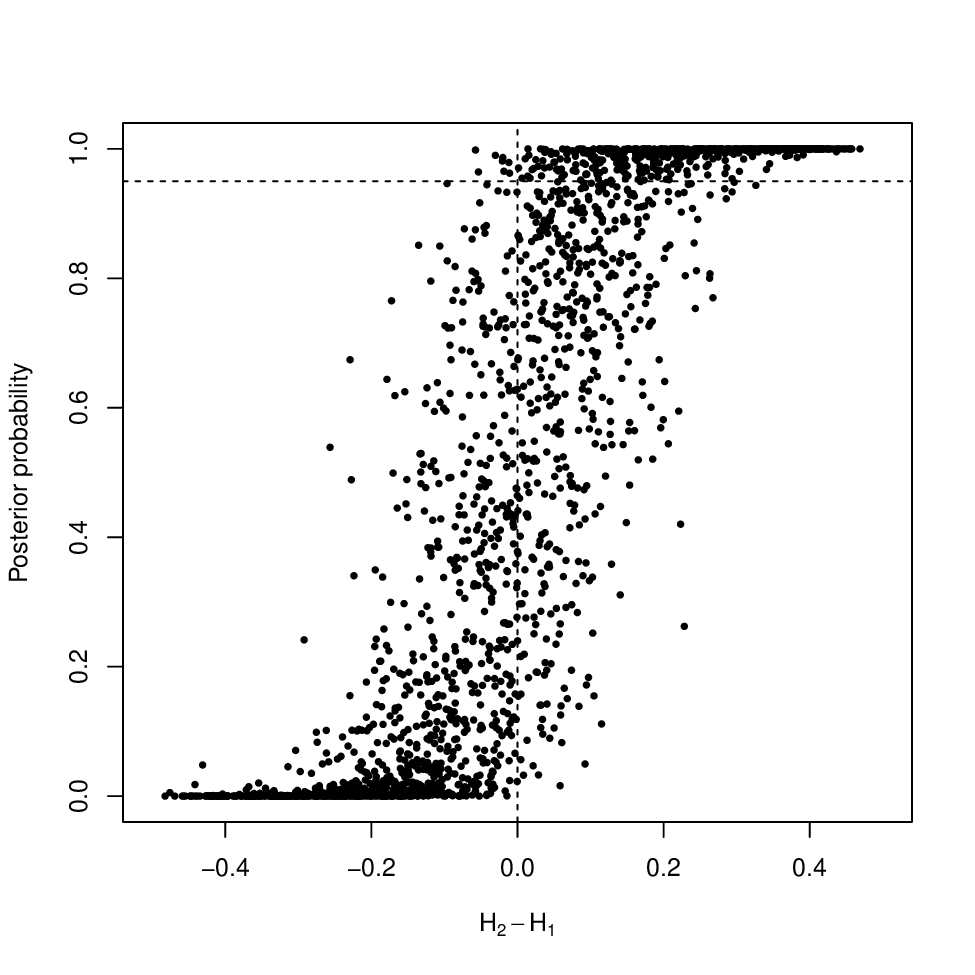} 
\includegraphics[width=0.32\textwidth]{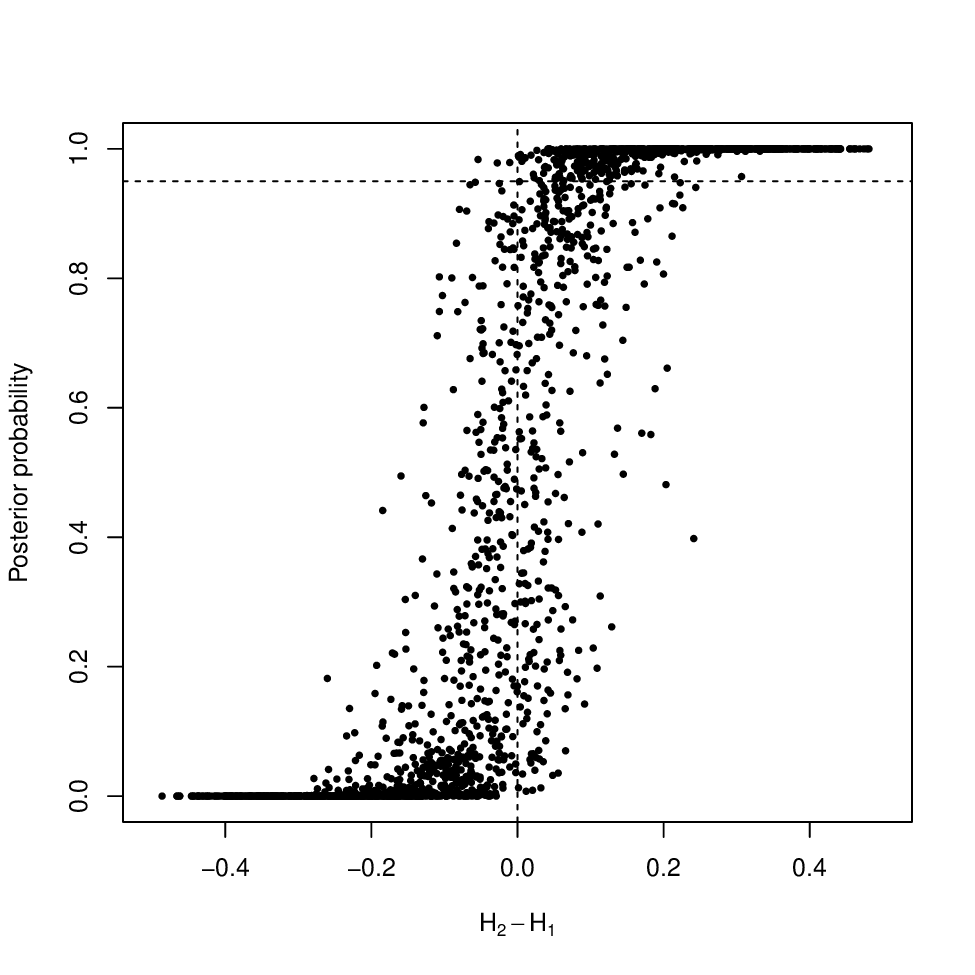}\\
\includegraphics[width=0.32\textwidth]{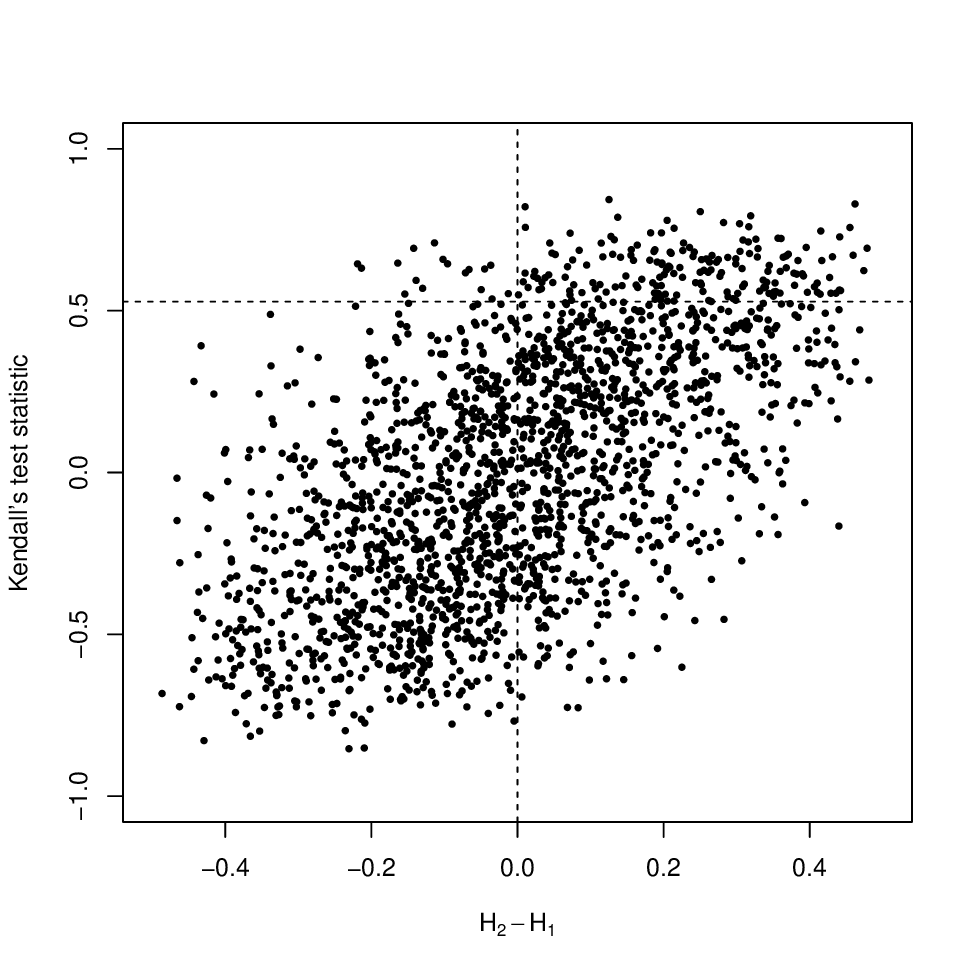} 
\includegraphics[width=0.32\textwidth]{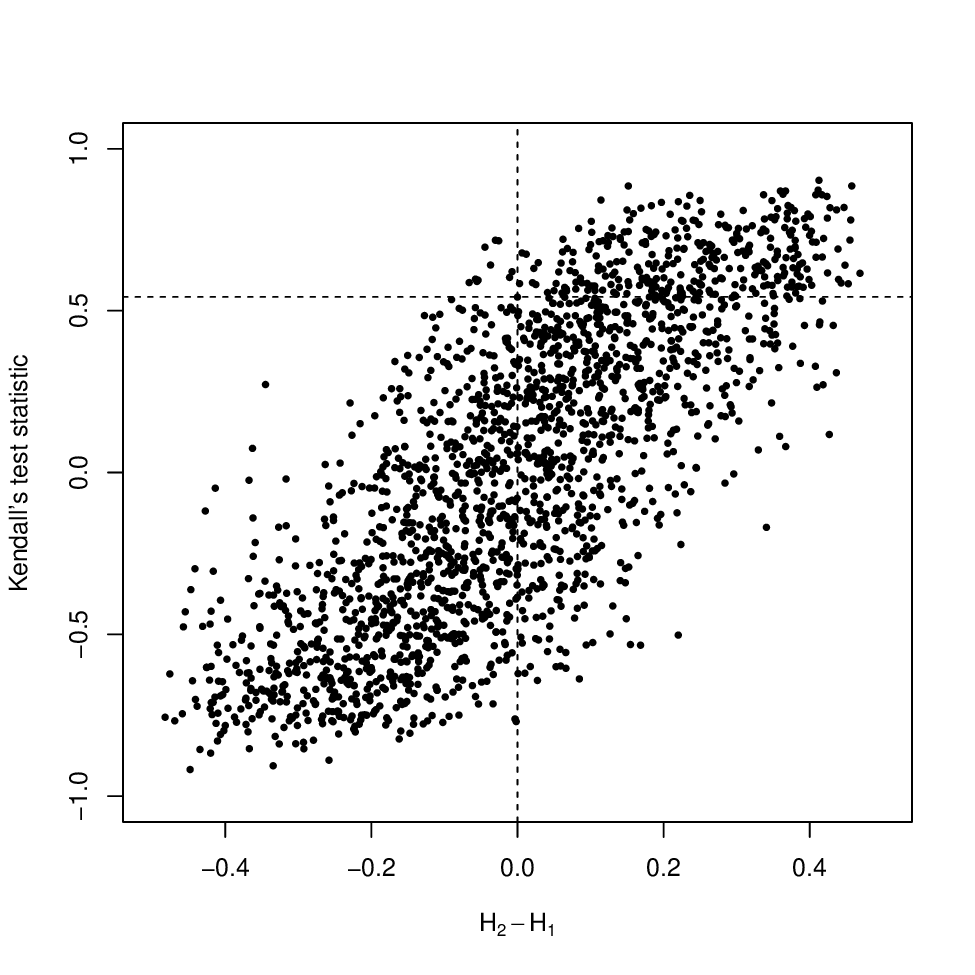} 
\includegraphics[width=0.32\textwidth]{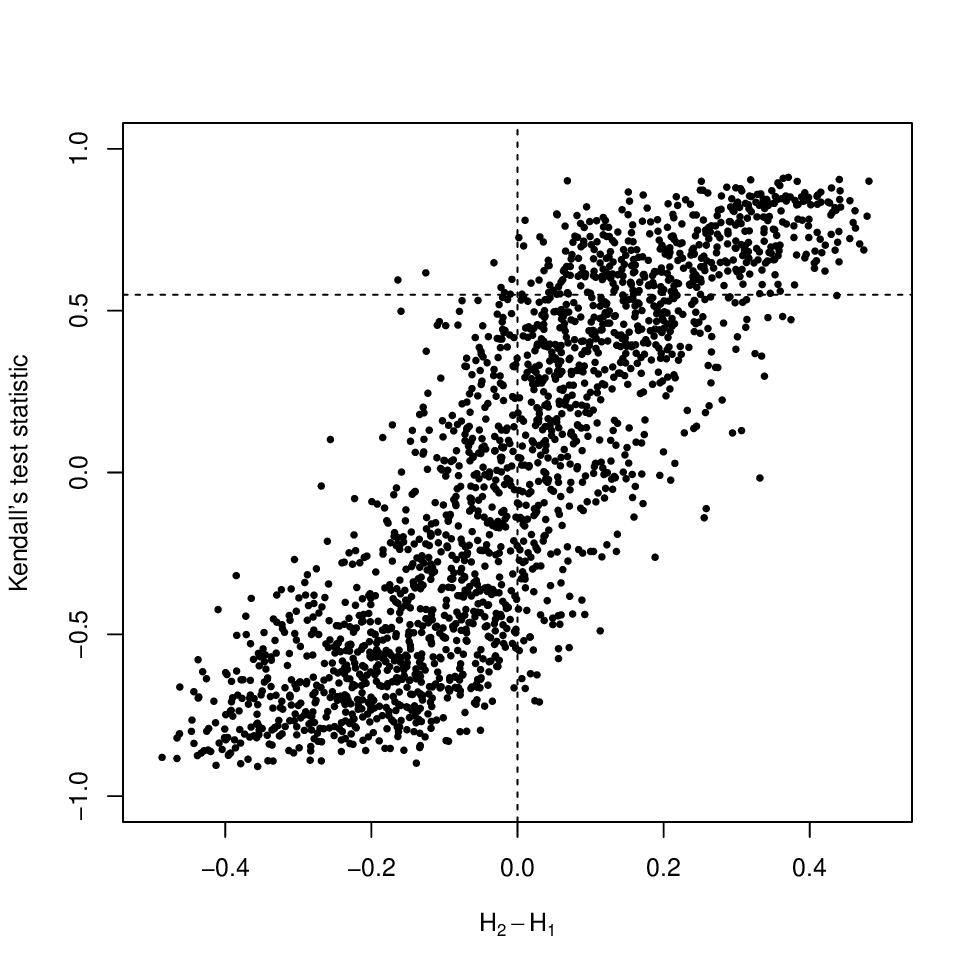}
\caption{Upper panels: Posterior mean estimates for the probabilities $\hat p = P(H_2-H_1>0\mid \mm{y})$ for time series generated by \eqref{eq:model} having length $n=200$ (left), $n = 500$ (middle) and $n = 1000$ (right).  The number of simulations for each $n$ is 2000 and the  horizontal lines correspond to $\hat p = 0.95$. Lower panels: Kendall's $\tau_K$ for local Hurst exponents of the same series, estimated in sliding windows of length $n/4$. The horizontal lines correspond to the $5\%$ upper quantiles of the test statistic $\tau_K$, assuming  a null hypothesis of $H_1 = H_2 = 0.75$.} 
\label{fig:posterior-prob}
\end{center}
\end{figure}

The upper panels of Figure~\ref{fig:posterior-prob} show the estimates of $\hat p$  as a function of the true differences $d = H_2-H_1$ for time series of lengths 200, 500, and 1000. As expected, these results demonstrate improved predictive performance in the form of reduced variability for larger $n$ and larger absolute differences $|H_2-H_1|$. For comparison, we also compute Kendall's $\tau$ statistic, 
$$\tau_K = \frac{\# \mbox{concordant pairs} - \# \mbox{disconcordant pairs}}{{p\choose 2}}$$
where $p=n-w$ is the number of valid window centres for sliding windows of length $w=n/4$. The resulting values of $\tau_K$ are shown in the lower panels of Figure \ref{fig:posterior-prob}, exhibiting substantially larger variability than the posterior means.

To further investigate these methods as early warning indicators of increased autocorrelation, we need to determine threshold values for declaring a positive detection. For the new model, we classify a series as exhibiting early warning signals when the posterior probability satisfies $\hat p>1-\alpha$, where $\alpha$ denotes a small probability. The horizontal lines in the upper panels of Figure~\ref{fig:posterior-prob} illustrate this threshold for $\alpha = 0.05$. Similarly, the horizontal lines in the lower panels correspond to the upper $5\%$ quantile for the distribution of $\tau_K$ under the null hypothesis of no change in the Hurst exponent. These quantiles were estimated by Monte Carlo simulation and are equal to $0.52$, $0.54$ and $0.55$, for $n=200, 500$, and $1000$, respectively.

\begin{table}[t]
\centering
\begin{tabular}{rrrrrrrrr}
\toprule
& \multicolumn{4}{c}{New model} & \multicolumn{4}{c}{Kendall's $\tau$ ($w = n/4$)} \\
$n$ & TPR & FPR  & PPV & NPV  &  TPR & FPR  & PPV & NPV \\
\midrule
200 & 0.294 & 0.009 & 0.970 &  0.597 &  0.182 & 0.017 & 0.912 & 0.559 \\
500 & 0.525 & 0.008 & 0.985 & 0.679 & 0.311 & 0.010 & 0.969 & 0.593 \\
1000 & 0.671 & 0.003 & 0.995 & 0.762 & 0.430 & 0.007  & 0.984 & 0.648  \\\hline
\end{tabular}
\caption{Performance measures predicting a positive increase in autocorrelation using the posterior mean $\hat p$ for the new model versus $\tau_K$ for running estimates in windows of length $n/4$. The  given measures include the true positive rate (TPR), the false positive rate (FPR), the positive predictive value (PPV) and the negative predictive value (NPV).}
\label{tab:measures}
\end{table}

The performance measures based on these thresholds are summarized in Table~\ref{tab:measures}, consistently showing improved classification performance using $\hat p$ versus $\tau_K$.  However, the true positive rate (sensitivity) for detecting an increase in  autocorrelation remains moderate. This can be explained by simulation choices, in which the difference $d = |H_2-H_1|$ is small in many cases, making detection challenging. In contrast, the false positive rate is very low, corresponding to a specificity exceeding $99\%$. A similar pattern appears in the positive and negative predictive values, which quantify the reliability of a positive or negative classification. Notably, the positive predictive values are very high, ensuring that a declared increase in autocorrelation is highly trustworthy.

\begin{figure}[h]
\begin{center}
\includegraphics[width=0.4\textwidth]{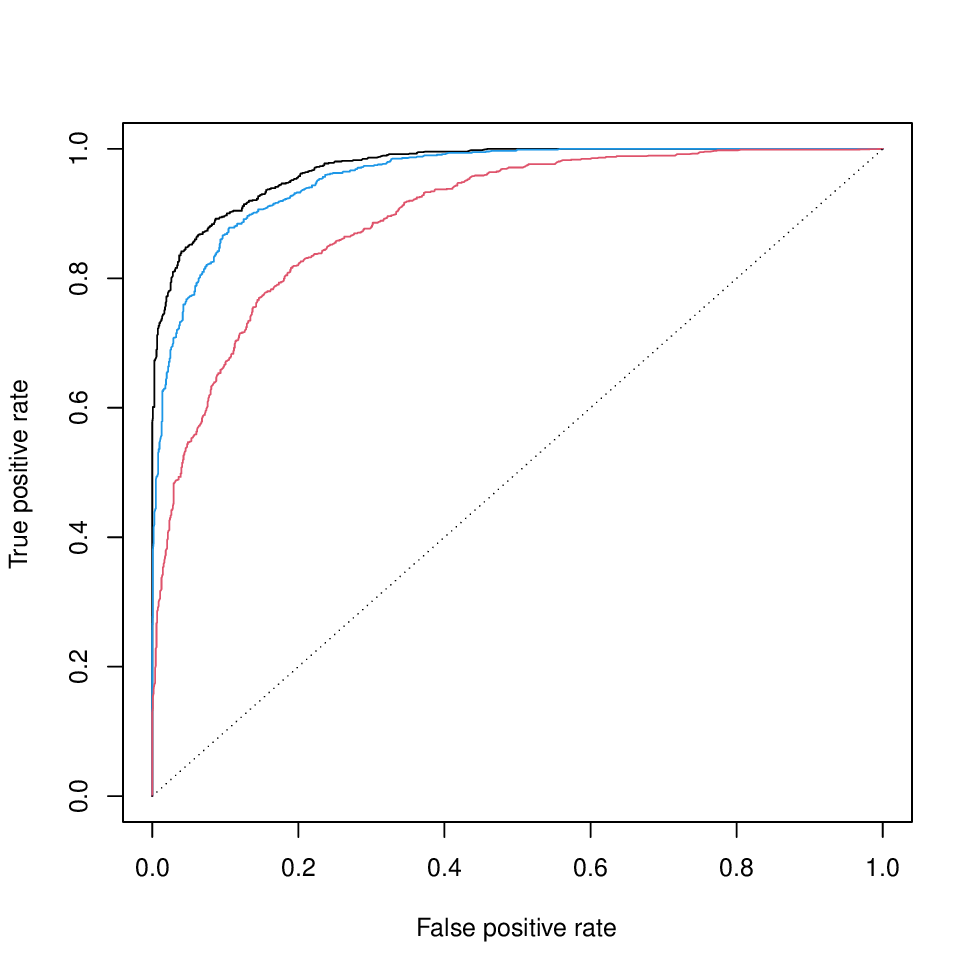}
\includegraphics[width=0.4\textwidth]{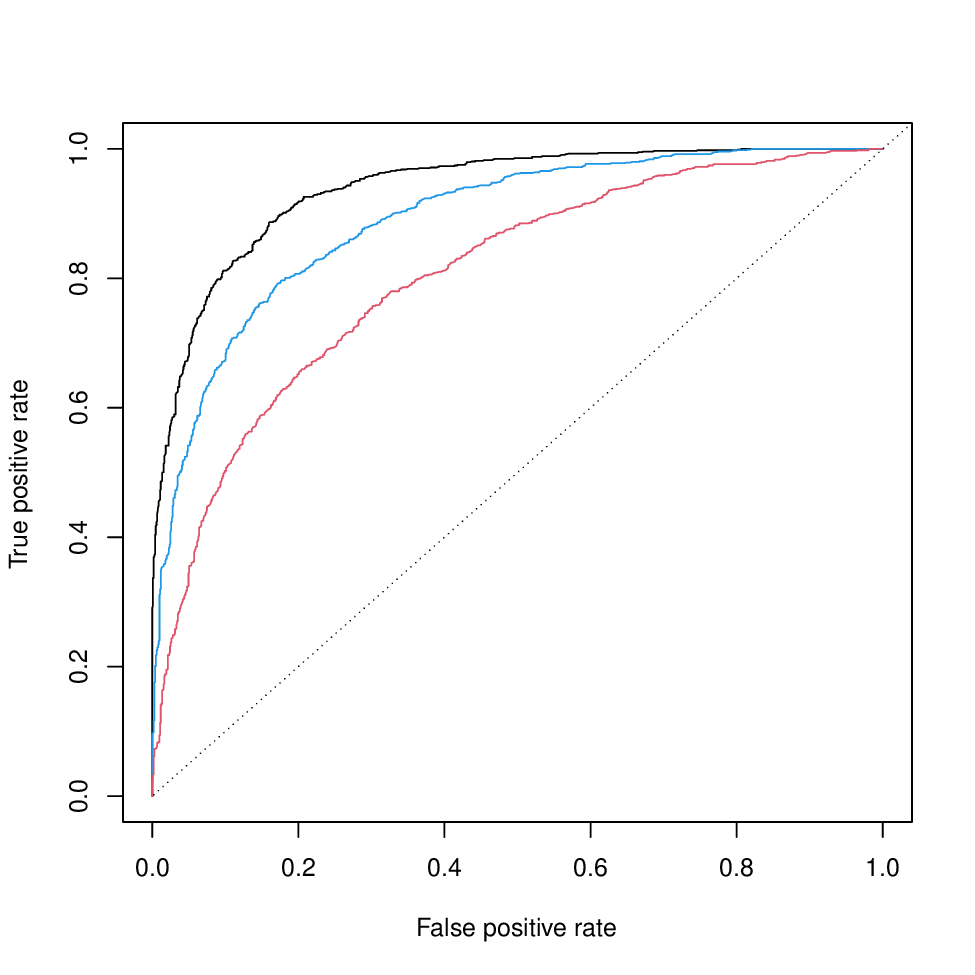}
\caption{Estimated ROC for the binary classification problem predicting the sign of $H_2 - H_1$  for time series of length $n=1000$ (black), $n=500$ (blue) and $n = 200$ (red). Classification using the posterior probabilities $\hat p$ (left) and $\tau_K$  for sliding window estimates of the Hurst exponent with window length $n/4$ (right).}
\label{fig:auc}
\end{center}
\end{figure}

The overall performance in classifying whether the autocorrelation is increasing or not is  illustrated by the ROC (Receiver Operating Characteristic) curves in Figure ~\ref{fig:auc}. These curves show the true positive rate (TPR) as function of the false positive rate (FPR) across all thresholds. For classification using the posterior probabilities $\hat p$, the area under the curve (AUC) increases from 0.895 for time series with $n=200$, to 0.957 and 0.972 for time series of lengths $n=500$ and $n = 1000$, respectively.  
 In comparison, the AUC for $\tau_K$ is estimated to be 0.803 for  $n = 200$. This increases to give AUC = 0.890 for $n = 500$ and further to 0.939 when $n = 1000$. These results were seen to be robust to the window length used (data not shown). 

\section{Early warning signals for climatic time series} \label{sec:4}
In this section, we fit the novel model to real climatic time series, including a reconstructed index for the variability of the North Atlantic sea surface temperatures \citep{michel:22}. We also investigate early warning signals for the paleoclimate records spanning the Dansgaard-Oeschger (DO) events \citep{dansgaard:84, dansgaard:93}. To analyse these events, the proposed model is modified to incorporate time-varying variance and adjusted to handle irregularly sampled observations. The results are compared with two previous studies using a Bayesian framework, but assuming short-memory dependence \citep{myrvoll1:25, myrvoll2:25}. 

 \subsection{Analysis for a reconstruction of the Atlantic multidecadal variability}
 Sea surface temperatures in the North Atlantic region are well-known to exhibit substantial multidecadal variability \citep{schlesinger:94, kushnir:94}. This phenomenon is commonly referred to as the Atlantic Multidecadal Variability (AMV), or Atlantic multidecadal oscillation \citep{wang:17, christiansen:25},  typically oscillating between warm and cold phases. The AMV has been shown to have coherent behaviour with the Atlantic Meridional Overturning Circulation (AMOC) \citep{zhang:17}, and a collapse of AMOC would have severe climatic impacts both in the Northern Hemisphere and globally. Such a collapse has been predicted to occur around the middle of this century \citep{ditlevsen:23}. In fact, it cannot be excluded that AMOC has already crossed a tipping point \citep{lenton:25}. However, these projections remain uncertain due to limited length of  observational records, biases in climate models, reliance on indirect proxies for AMOC strength, and simplifying model assumptions \citep{loriani:25, lenton:25}. Direct observations of the natural variability of  the AMV are also limited, only including approximately the past 150 years, a period that also coincides with substantial  anthropogenic climate forcing.

\begin{figure}[h]
\center
\includegraphics[width=0.5\textwidth]{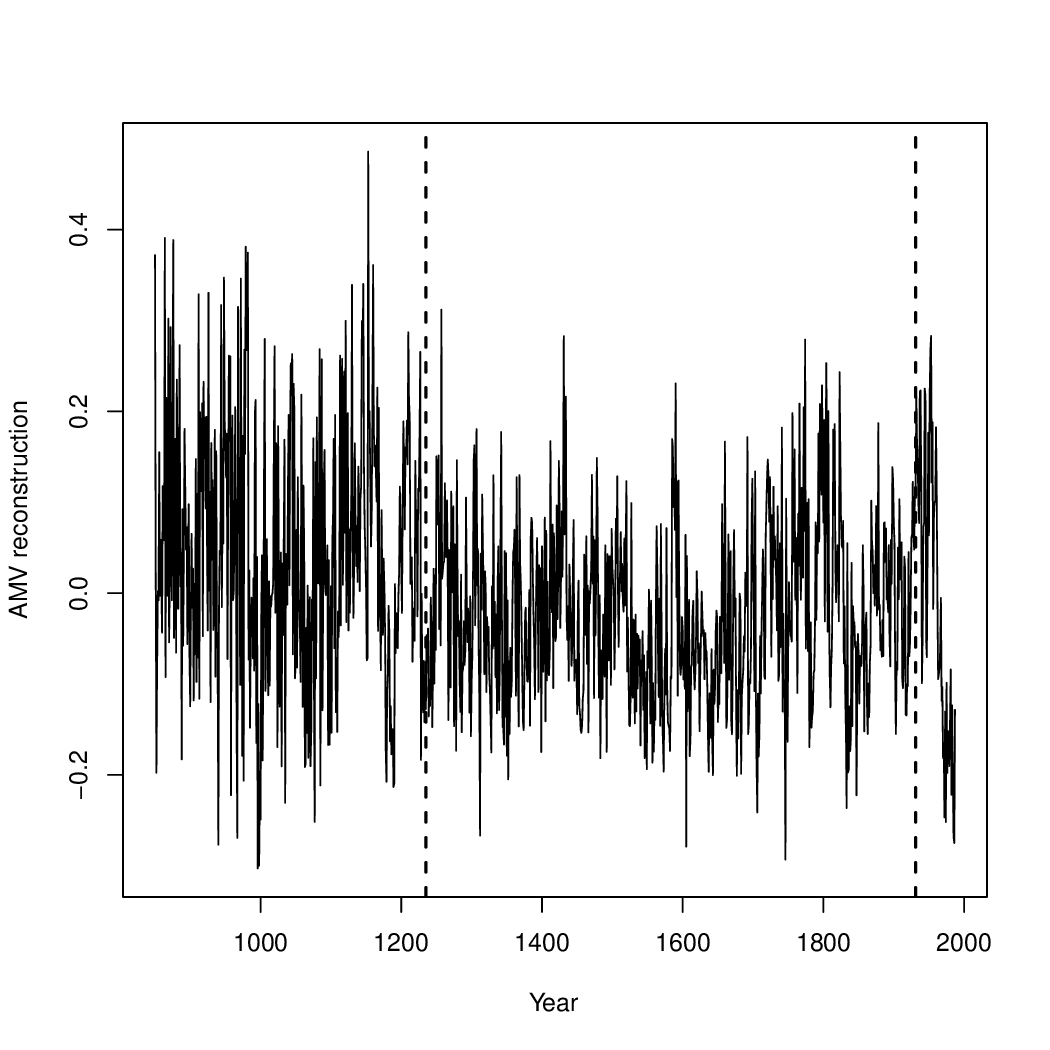}
\caption{The reconstructed AMV given by \citet{michel:22} for the period 850 - 1987. The vertical lines denote changepoints in variance and the series analysed correspond to the time period of 1235 - 1931. }
\label{fig:amv}
\end{figure}

Here, we analyse a reconstruction of the AMV over the last millennium provided by \citet{michel:22}, shown in Figure \ref{fig:amv}. Fitting \eqref{eq:approx-model} to the series shows a large increase in Hurst exponents, giving $H_1 = 0.61$ and $H_2 = 0.92$. However, the variance of this series clearly changes in time. The two dashed vertical lines illustrate changes in variance around the years 1235 and 1931, identified by change point analysis \citep{killick:14}. To focus on more subtle changes in the autocorrelation, we fit  \eqref{eq:approx-model} to the segment of the series between these two change points, yielding a time series of length $n = 697$ years. Detrended fluctuation analysis \citep{mesquita:20} shows that this segment exhibits long-range dependence, estimating a power-law relationship with Hurst exponent $\hat H = 0.781$. 

\begin{figure}[h]
\begin{center}
\includegraphics[width=0.32\textwidth]{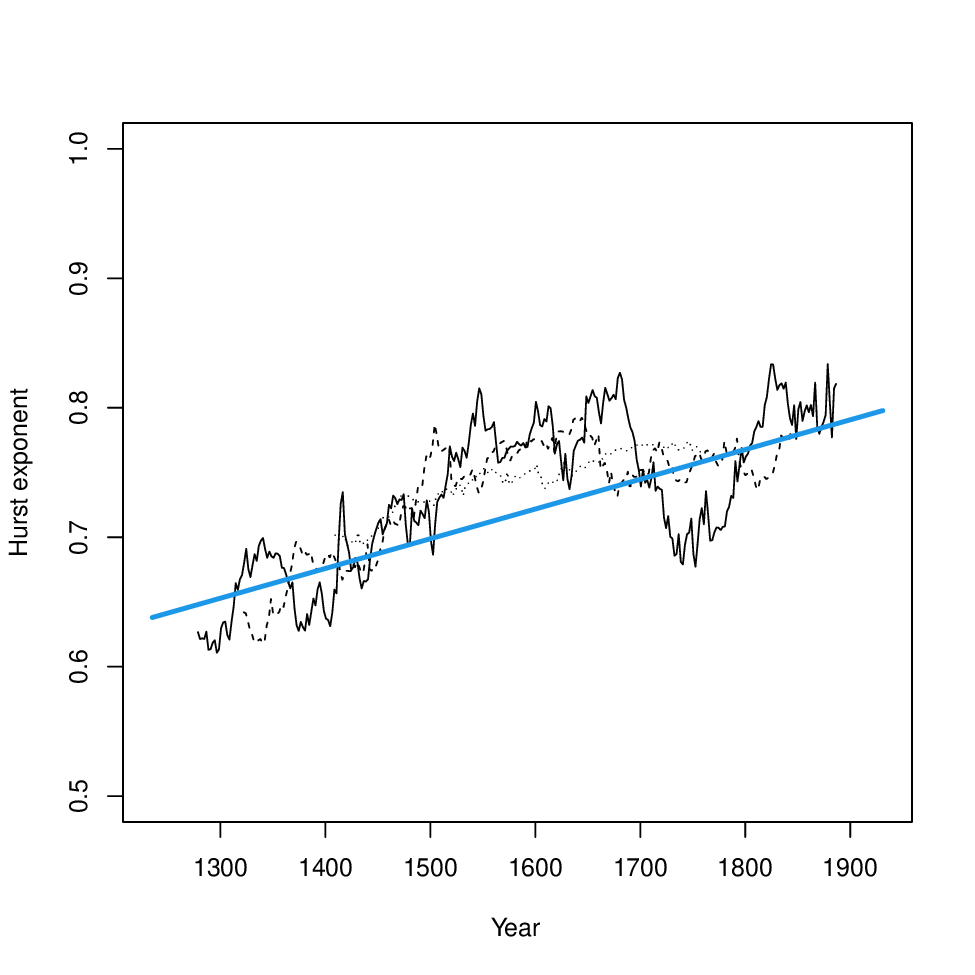}
\includegraphics[width=0.32\textwidth]{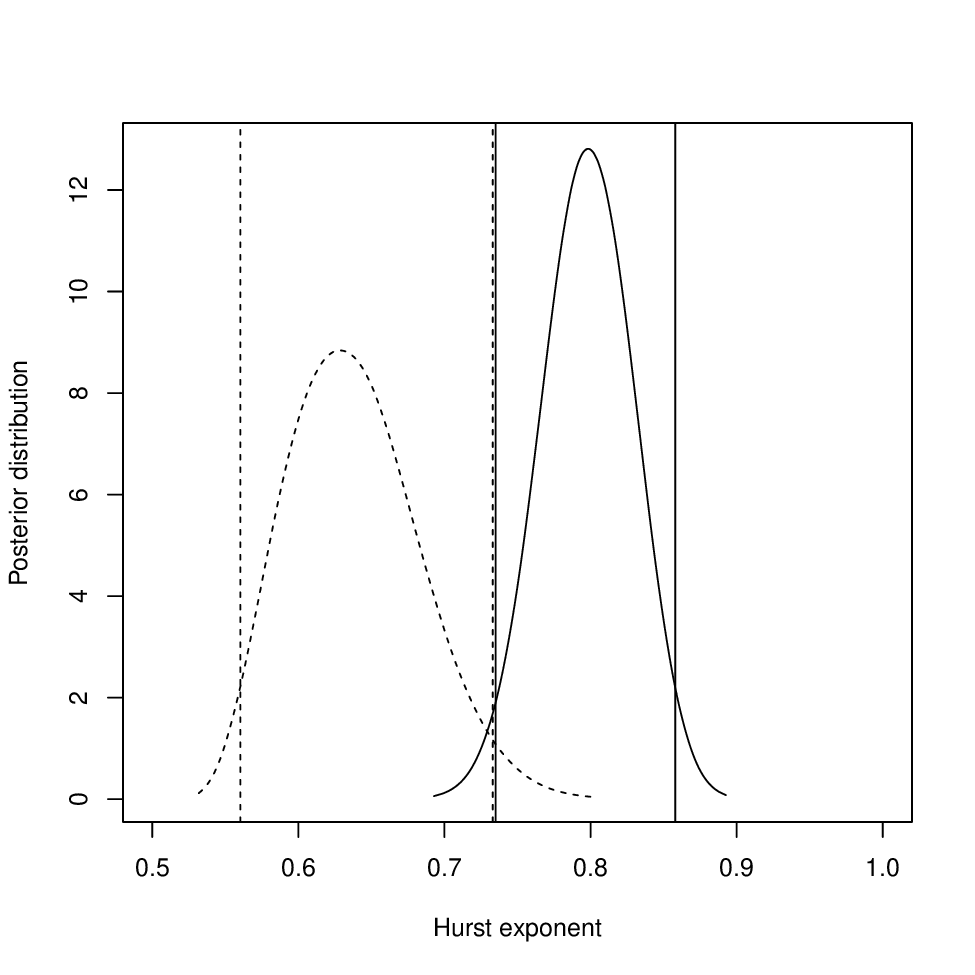}
\includegraphics[width=0.32\textwidth]{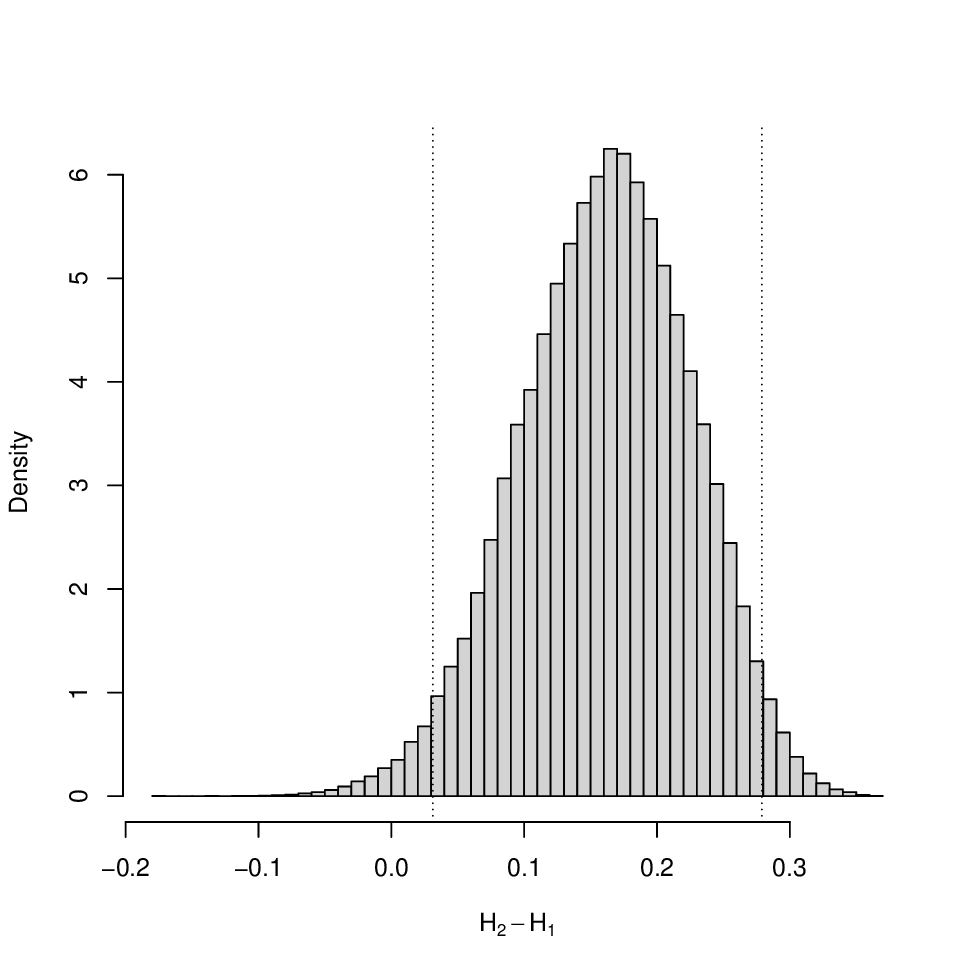}
\caption{Left: The estimated linear increase in the Hurst exponent (blue) and running estimates using window sizes of length $n/8$ (solid), $n/4$ (dashed) and $n/2$ (dotted). Middle: The estimated marginal posterior distributions for $H_1$ and $H_2$ including $95\%$ credible intervals.  Right: The approximated posterior distribution of the difference $H_2-H_1$ including $95\%$ confidence interval (dotted vertical lines).}
\label{fig:running-estimates}
\end{center}
\end{figure}

Fitting \eqref{eq:approx-model} to the segment yields a posterior mean estimate of $H_1$ equal to 0.638, with a $95\%$ credible interval of $(0.560, 0.733)$. The corresponding results for $H_2$ give posterior mean equal to 0.798, with a $95\%$ credible interval of $(0.735, 0.858)$  The estimated increase in $H$ is illustrated in Figure~\ref{fig:running-estimates}. This is seen to match well with running estimates of the Hurst exponents, computed  for window lengths $n/8$, $n/4$ and $n/2$.  The figure also displays the estimated posterior marginal distributions of $H_1$ and $H_2$, as well as the posterior distribution of  the difference $H_2-H_1$. The latter is approximated using $B =100{,}000$ Monte Carlo samples from the joint posterior distribution of the hyperparameters. The estimated posterior probability of a positive difference is $\hat p = \hat P(H_2-H_1>0 \mid \mm{y}) =0.992$, and the corresponding $95\%$ credible interval for $H_2-H_1$ is $(0.031, 0.279)$. 

In conclusion, the results indicate an increase in the Hurst exponent for the given segment of the original time series, and also for the full series. This suggests increasing  autocorrelation in the series, consistent with the findings reported by  \citet{michel:22}. 

\subsection{Analysis of the Dansgaard-Oeschger events}\label{sec:4.2}
\subsubsection{The data set}
The Dansgaard-Oeschger (D-O) events refer to the abrupt climate fluctuations occurring during the last glacial period  \citep{dansgaard:84, dansgaard:93}. Based on proxy records  from Greenland ice cores, the D-O events are characterised by colder stadial phases, followed by rapidly increasing temperatures during interstadial phases. These events have been considered to provide critical insight into drivers of abrupt climate changes and potential early warning signals of the stadial time series have been studied extensively using a variety of methods \citep{dakos:08, ditlevsen:10, lenton:13, rypdal:16, boers:18, lohmann:19, myrvoll1:25, myrvoll2:25, hummel:25}. However, conclusions remain deeply divided.
\begin{figure}[h]
\includegraphics[width=0.99\textwidth]{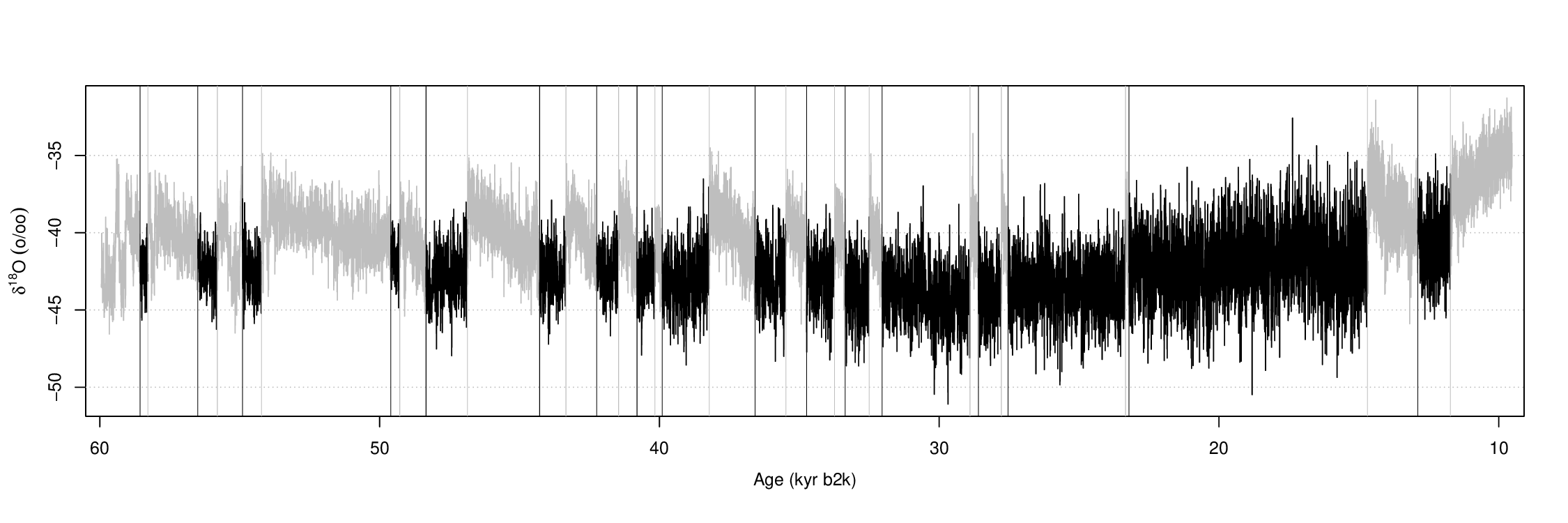}
\caption{The $\delta^{18} $O proxy records from the NGRIP core, dating back to 60.2 ka b2k. }
\label{fig:raw-ews}
\end{figure}

We analyse the D-O events using $\delta^{18}$O  
proxy records from the NGRIP ice core (North Greenland Ice Core Project, \citet{ngrip:04, gkinis:14}).  The data are sampled at a regular 5cm depth resolution. Using the Greenland Ice Core Chronology 2005 (GICC05) timescale, the measurements  correspond to an irregularly sampled time series extending back to 60.2 ka b2k \citep{vinther:06, rasmussen:06, andersen:06, svensson:08}. The full record is shown in Figure \ref{fig:raw-ews}. The time series analysed here correspond to cold stadial phases (black), which precede interstadial phases characterised by initial abrupt warming (grey). In total, the record contains 17 transitions from cold to warm conditions. Some stadial segments contain relatively few observations, rendering inference highly uncertain.

\subsubsection{Incorporating time-varying variance}\label{sec:4.2.2}
Early warning signals preceding bifurcation points are typically characterised by gradual increases in both variance and autocorrelation. To accommodate changes to the marginal variance, the proposed model is extended to   
\begin{equation}
\tilde{\epsilon}(t_i) = \sigma(t_i) \cdot \tau^{-1/2}\left(\sqrt{1-w(t_i)}x_{H_1}(t_i) + \sqrt{w(t_i)}x_{H_2}(t_i)\right) \label{eq:model-sigma}
\end{equation}
where the  parametric function $\sigma(t_i)$ represents deviation from the overall standard deviation level $\tau^{-1/2}$. We model the time-varying standard deviation by  a logistic function 
\begin{equation}
  \sigma(t_i) = \frac{1}{2} + \frac{1}{1+e^{-\beta\left(\frac{t_i-t_1}{t_n-t_1}-\frac{1}{2}\right)}},  
  \label{eq:logitvariance}
\end{equation}
which includes a hyperparameter $\beta$ to be estimated. This specification allows a weakly time-varying standard deviation with range $0.5\tau^{-1/2}$ to $1.5\tau^{-1/2}$, reducing to 1 when $\beta=0$. The hyperparameter $\beta$ is assigned a Laplace prior distribution with scale parameter 1, inducing shrinkage to 0. Alternative parametric forms for the time-varying standard deviation can readily be adopted if required by a specific application.  

\subsubsection{Regression model and results}
The 17 time series are first standardized to ensure consistent interpretation of the prior distributions for the hyperparameters. The expected values of the series, $\eta(t_i) = E(y(t_i))$, is then modelled by the additive predictor
\begin{equation}
\eta(t_i) = \mu(t_i) + \tilde{\epsilon}(t_i), \label{eq:regr-model}
\end{equation}
in which $\mu(.)$ is assigned a scaled intrinsic GMRF prior distribution \citep{rue:05, sorbye:14},  while $\tilde\epsilon(.)$ represents the new model component in \eqref{eq:model-sigma}. The model is implemented for irregularly sampled time series as explained in Appendix \ref{app:irregular}. 

The posterior mean estimates of the hyperparameters $H_1$, $H_2$, and $\beta$ are displayed in Table~\ref{tab:do-events}, also including the probability of a positive difference between the two Hurst exponents, $\hat p = P(H_2-H_1>0\mid \mm{y})$. 
\begin{table}[ht]
\centering
\begin{tabular}{rrrrrrrr}
\toprule
no & from & to  &  $n$ &  $H_1$ & $H_2$  & $\beta$ & $\hat p$ \\
\midrule
1 & 12896 & 11728 & 667 &  0.962 &  0.948  & -0.007 & 0.118  \\
2 & 23220 & 14692 & 3713 &  0.906 & 0.913 & 0.512  &  0.633  \\
3 & 27540 & 23340 & 1370 & 0.789 & 0.841 &  0.037 & 0.897 \\
4 & 28600 & 27780 & 270 & 0.744 & 0.758 &  -0.642 & 0.562\\
5 & 32040 & 28900 & 962 & 0.855 & 0.806 & 0.424 & 0.195 \\
6 & 33360 & 32500 & 252 & 0.793 & 0.825 &  -0.031 & 0.683  \\
7 & 34740 & 33740 & 315 & 0.777 & 0.805 & 0.172& 0.622 \\
8 & 36580 & 35480 & 345 & 0.808 & 0.825 &  0.430& 0.556 \\
9 & 39900 & 38220 & 492 & 0.833 & 0.792 & 0.597 & 0.296 \\
10 & 40800 & 40160 & 202 & 0.781 & 0.723 &  -0.280 & 0.319 \\ 
11  & 42240 & 41460  & 219 & 0.651 & 0.697&  0.861 & 0.682   \\ 
12 & 44280 & 43340 & 251 & 0.775 & 0.859 &  -0.462 & 0.775\\
13 & 48340 & 46860 & 373 & 0.844 & 0.792 &  0.399 & 0.246  \\
14 & 49600 & 49280 & 91 & 0.639 & 0.666 &  0.188 & 0.514 \\
15 & 54900 & 54220 & 163 & 0.724 & 0.704&  -0.726& 0.443  \\
16 & 56500 & 55800 & 162 & 0.903 & 0.834 &  0.780 & 0.166  \\
17 & 58560 & 58280 & 70 & 0.738 & 0.759 &  0.032 & 0.719 \\\hline 
\end{tabular}
\caption{Posterior mean estimates for hyperparameters, analysing the 17 time series representing the  cold stadial proxy data from NGRIP. The table also includes the posterior probability of increased Hurst parameter $\hat p = P(H_2-H_1>0\mid \mm{y})$.
 The series are numbered from 1 - 17 in reversed chronological order. \label{tab:do-events}}
\end{table}

Based on the estimated Hurst exponents and additional analyses using detrended fluctuation analysis (not shown), the assumption of a long-range dependency structure is justified. 
The results provide no clear evidence of significant increases in the Hurst exponents, with $\hat p <0.95$ in all cases. This is not surprising, especially as many of the series are rather short and probably subject to observational noise. Also, this finding is consistent with the conclusions of the \citet{ditlevsen:10} and \citet{hummel:25}, who likewise found no systematic evidence of early warning signals in these records. 

By contrast, several studies have reported significant findings of early warning signals for some of the D-O events, including \citet{rypdal:16}, \citet{boers:18}, \citet{myrvoll1:25} and \citet{myrvoll2:25}.  Assuming short-memory dependence, the latter two studies reported a significant increase in lag-one autocorrelation for the periods 2, 5, 9, 11 and 16, using a Bayesian framework similar to ours. 
However, a limitation of these two studies is that they rely on standard CSD theory, assuming that the time series satisfy the Langevin differential equation 
\begin{equation}
dx(t) = -\lambda(x(t)-x_s)dt + \sigma dB(t).\label{eq:langevin}
\end{equation}
This formulation combines deterministic and stochastic forcing. The deterministic component represents internal system dynamics and is governed by the control parameter $\lambda$, which determines the recovery or mean-reversion rate toward the stable fixed point $x_s$. The parameter $\sigma$ scales the stochastic forcing induced by the Brownian motion $B_t$. 
The solution to this equation is an Ornstein-Uhlenbeck process, which, after discretization, corresponds to  an AR(1) process with lag-one autocorrelation $\phi = e^{-\lambda\Delta t}$ and  marginal variance $\sigma^2(1-\phi^2)/(2\lambda)$.  According to the CSD theory near bifurcation points, the control parameter $\lambda$ is expected to approach 0, implying increased variance and a lag-one autocorrelation coefficient $\phi$ approaching unity, see e.g.  \citet{scheffer:09}. A potential limitation of this formulation is that $\lambda$ appears in both the variance and the autocorrelation, implying that changes in variance and dependence structure may be confounded with amplification of the external stochastic forcing, in which $\sigma$ would depend on time \citep{boers:21}.

The models used in \citet{myrvoll1:25} and \citet{myrvoll2:25} assume a common dependence on $\lambda$ and constant $\sigma$, assumptions which do not appear to hold for the D-O events. These models are then prone to falsely detecting increased autocorrelation for time series where changes are driven solely by increased marginal variance. Further elaborations are given in Appendix \ref{app:ar1-results}, which also includes a modified version of the method in \citet{myrvoll1:25} including a time-varying standard deviation as defined in \eqref{eq:logitvariance}. Using this modified approach, the results are consistent with ours, showing no significant increases in autocorrelation for these series, except for a spurious detection in no. 17 due to the short length of 70 observations.

\section{Discussion and future work} \label{sec:5}
Increases in the variance and autocorrelation of time series  represent important and commonly used statistical early warning signals of critical transitions. These indicators are well-motivated by theory of CSD, describing the reduced ability of a dynamical system to recover from small perturbations as it approaches a bifurcation point. For short-term memory processes, the lag-one autocorrelation represents a natural indicator for changes in the dependency structure.
However, this indicator might be sensitive to amplified external noise, either masking true internal instability or giving spurious early warning signals \citep{boers:21, dakos:12, lenton:12}. Another limitation is the validity of this indicator when short-term memory assumptions are violated.

Many climatic time series exhibit long-range dependence, having  power-law correlations consistent with fGn or fractional Brownian motion \citep{koscielny-bunde:98, rybski:08,  lovejoy:13, rypdal:16,  wang:16}. Defining an fGn model with time-varying Hurst exponent  $H(t)$ is not straightforward, neither conceptually nor computationally. From a conceptual perspective, the Hurst exponent characterises dependence across all time scales. Treating it as a local, time-varying quantity, raises questions regarding interpretation and identifiability.  However, an increase in local Hurst exponents can be seen as corresponding to enhanced persistence and longer correlation times, consistent with loss of stability near critical transitions.  From a computational perspective, existing model formulations typically require non-stationary covariance structures or stochastic integration with respect to fractional Brownian motion \citep{decreusefond:05, ryvkina:15}, for which multiple, non-equivalent definitions exist.

The approach proposed in this paper circumvents these difficulties by representing time-varying dependence through a mixture of two fractional Gaussian noise processes with a time-dependent weight function. This construction yields an effective approximation to a long-range dependent process with evolving memory properties, while retaining a stationary Gaussian structure for fixed weights. By mapping the weights to time-varying Hurst exponents, these play an analogous role to lag-one autocorrelations in short-memory time series. 

As a result, the model remains conceptually transparent and computationally efficient, and it can be implemented within the latent Gaussian modelling framework using \texttt{R-INLA}. 
Importantly, this gives a model-based approach to detect increased autocorrelation beyond the first lag, also allowing for joint estimation with other latent model components. This eliminates a preliminary detrending step which could introduce biased results and ensures proper control of the model components and their interaction.  Also, the Bayesian framework provides a rigorous approach to uncertainty quantification of all parameter estimates and the use of priors with proper shrinkage, like penalised complexity priors, ensures stable and robust inference. Furthermore, a model-based approach avoids the use of sliding  window estimates which are sensitive to the choice of window size. 

In general, detecting early warning signals in real-world climatic data remains challenging. Many real time series have limited record lengths combined with measurement error, constraining validity of the conclusions made.
The proposed Bayesian framework provides a flexible basis for addressing these challenges, but it does not eliminate the fundamental uncertainty associated with inference near potential tipping points. Different modelling approaches and assumptions can lead to deeply divergent conclusions. This is exemplified by the records spanning the D-O events. Despite decades of research, there is still no consensus on the presence or reliability of early warning signals in these paleoclimate time series \citep{hummel:25}. Our conclusions of no significant early warning signals for the D-O records, is supported by previous studies \citep{ditlevsen:10, hummel:25} and also by renewed analysis using a modification of the approach in  \citet{myrvoll1:25}. Especially, an assumption of joint increase in the variance and autocorrelation of these series, does not seem to be correct. Using the proposed model, we find it important to model a time-varying variance independently from the autocorrelation. 

An important extension of the proposed model is to include climate forcing. This is especially relevant analysing more recent climatic records, where anthropogenic forcing often plays a dominant role. The fGn representation using a weighted sum of AR(1) processes, closely resembles multibox energy balance models \citep{fredriksen:17}, only differing in how the weights and lag-one autocorrelation coefficients are optimized. This connection suggests a natural extension in which forcing terms are added to the Langevin equation, incorporated for a sum of AR(1) processes. This would allow the forcing response for different time scales to be studied explicitly, representing an interesting future direction of research.   

\section{Appendix}\label{sec:appendix}
\subsection{Asymptotic autocorrelation for the mixed model component}\label{app:asymptotic}
Asymptotically, the autocovariance function of the new mixed model has the same properties as for a stationary fGn. For simplicity, assume that $H_2>H_1$ and let $w = w(t)$.  Then 
\begin{eqnarray*}
\lim_{k\rightarrow \infty}\gamma_{\epsilon}(k) &\sim & (1-w) H_1(2H_1-1)k^{2(H_1-1)} + w  H_2(2H_2-1)k^{2(H_2-1)} \\
& = & w H_2(2H_2-1)k^{2(H_2-1)}\left(1+\frac{(1-w)H_1(2H_1-1)k^{2(H_1-1)}}{w H_2(2H_2-1)k^{2(H_2-1)}}\right)\\
& \sim & w H_2(2H_2-1)k^{2(H_2-1)}\left(1+ k^{2(H_1-H_2)}\right)\\
& \sim &  wH_2(2H_2-1)k^{2(H_2-1)}
\end{eqnarray*}
As $k\rightarrow \infty$, the weight $w=w(t)$  will be 1 and  the asymptotic properties are similar to a stationary fGn process with Hurst exponent $H_2$. 

\subsection{Implementation within the class of latent Gaussian models}\label{app:implementation}
The model component $\tilde{\mm{\epsilon}} = (\epsilon(t_1),\ldots , \epsilon(t_n))^\top$ defined by \eqref{eq:ar-approx} and \eqref{eq:approx-model} can be implemented within the class of latent Gaussian models  using the \texttt{rgeneric} tool available in the \texttt{R-INLA} package. 

First, we need to derive the  precision matrix for the stacked vector of $\tilde{\mm{\epsilon}}$ and its defining components, conditioned on hyperparameters $\mm{\theta}$. Define the stacked vector by 
\begin{equation*}
    \mm{u} = (\tilde{\mm{\epsilon}}^\top, \mm{Z}_1^\top,\mm{Z}_2^\top)^\top ,
\end{equation*}
 where
\begin{equation*}
    \boldsymbol{Z}_i = (\mm{z}_{i1}^\top, \mm{z}_{i2}^\top,\ldots, \mm{z}_{im}^\top)^\top, \qquad i=1,2
\end{equation*}
denote the vectors containing the $m$ AR(1) processes used to approximate the fGn $\mm{x}_{H_i}$. As the AR(1) processes are assumed to be independent, the precision matrix of $\boldsymbol{Z}_i$ is given by the block diagonal matrix
\begin{equation*}
    \boldsymbol{R}(H_i) = \text{diag}\left(
        \boldsymbol{Q}_{\phi_1(H_i)}, \boldsymbol{Q}_{\phi_2(H_i)}, \ldots,\boldsymbol{Q}_{\phi_m(H_i)} \right),
\end{equation*}
where for $j = 1,\ldots , m$
\begin{equation*}
    \boldsymbol{Q}_{\phi_j(H_i)}= \frac{1}{1-\phi_j^2}\begin{pmatrix}
        1 & -\phi_j(H_i) &&& \\
        -\phi_j(H_i) &1+\phi_j^2(H_i)  & -\phi_j(H_i) &&\\
         & & & &\\
        & \ddots &\ddots &\ddots & \\
          & & & &\\
        &&-\phi_j(H_i) & 1+\phi_j^2(H_i) & -\phi_j(H_i) \\
        &&& -\phi_j(H_i) & 1
    \end{pmatrix}.
\end{equation*}
Specifically, each  AR(1) process $\mm{z}_{ij}$  has a fixed lag-one autocorrelation coefficient  $\phi_j(H_i)$ approximating the fGn $\mm{x}_{H_i}$. 
The approximations for an fGn is uniquely defined for all Hurst exponents in the interval $(0.51,0.99)$  and 
 readily available using the sum of  $m=3$ or  $m=4$ AR(1) processes \citep{sorbye:19}. Here, we have chosen to use $m=4$ as this gives a slightly more accurate approximation than $m=3$. 
  
In matrix notation, the proposed  model can now be expressed by  
\begin{equation}
    \tilde{\mm{\epsilon}} = \tau^{-1/2}(\boldsymbol{A}_1 \boldsymbol{Z}_1 +\boldsymbol{A}_2 \boldsymbol{Z}_2) + \kappa, \label{eq:model-mat}
\end{equation}
where
\begin{eqnarray*}
    \boldsymbol{A}_i &= &\begin{pmatrix}
        \sqrt{v_{i1}}\boldsymbol{D}_i , \sqrt{v_{i2}}\boldsymbol{D}_i,\ldots,\sqrt{v_{im}}\boldsymbol{D}_i
    \end{pmatrix}. 
    \end{eqnarray*}
This includes the fixed set of weight $\{v_{ik}\}$ for each approximate fGn and the 
diagonal matrices 
$$
       \boldsymbol{D}= \text{diag}(\sqrt{1-\boldsymbol{w}}),\quad \boldsymbol{D} = \text{diag}(\sqrt{\boldsymbol{w}}).
$$
including the  weight function $\mm{w} = (w(t_1),\ldots , w(t_n))^{\top}$ of the proposed model. 
In \eqref{eq:model-mat},  we have added a small Gaussian noise term $\kappa\sim \pi_G(0, \tau^{-1}_h)$
with a high fixed precision $\tau_h = \exp(15)$ 
to ensure a non-singular distribution.

To derive the precision matrix of the stacked vector $\mm{u}$, we first express its conditional distribution given the hyperparameters by    
\begin{equation*}
     \pi(\mm{u}\mid \mm{\theta})= \pi(\tilde{\mm{\epsilon}}\mid \boldsymbol{Z}_1, \boldsymbol{Z}_2,\boldsymbol{\theta})\pi(\boldsymbol{Z}_1,\boldsymbol{\theta})\pi(\boldsymbol{Z}_2,\boldsymbol{\theta})
\end{equation*}
where
\begin{eqnarray*}
    \tilde{\mm{\epsilon}} \mid \mm{Z}_1, \mm{Z}_2, \mm{\theta} &\sim & 
    \pi_G\left(\tau^{-1/2} (\mm{A}_1 \mm{Z}_1 + \mm{A}_2 \mm{Z}_2), \tau_h^{-1}\right) \\
    \mm{Z}_i  &\sim &  \pi_G \left( \mm{0}, \mm{R}_i (H_i)\right), \quad i=1,2
\end{eqnarray*}
Focusing on the exponent of the resulting Gaussian distribution, we get
\begin{eqnarray*}
    \pi(\mm{u}\mid \boldsymbol{\theta}) &\propto &\exp\left\{ -\frac{1}{2} (\tilde{\mm{\epsilon}} - \tau^{-1/2}(\boldsymbol{A}_1\boldsymbol{Z}_1+\boldsymbol{A}_2\boldsymbol{Z}_2))^\top \tau_h
    \boldsymbol{I} (\tilde{\mm{\epsilon}}- \tau^{-1/2}(\boldsymbol{A}_1\boldsymbol{Z}_1+\boldsymbol{A}_2\boldsymbol{Z}_2))\right\}\\
  && \cdot \exp\left\{-\frac{1}{2}\left(\boldsymbol{Z}_1^\top \boldsymbol{R}(H_1)\boldsymbol{Z}_1 + \boldsymbol{Z}_2 ^\top\boldsymbol{R}(H_2)\boldsymbol{Z}_2\right)\right\}
\end{eqnarray*}
Further expansion and reordering yields
\begin{eqnarray*}
    \pi(\boldsymbol{u}\mid \boldsymbol{\theta}) & \propto&  \exp {\bigg\{} -\frac{1}{2}{\bigg(} 
    \tau_h \tilde{\mm{\epsilon}}^\top \boldsymbol{I}\tilde{\mm{\epsilon}}  - 2 \tau^{-1/2} \tau_h \tilde{\mm{\epsilon}}^\top ( \boldsymbol{A}_1\boldsymbol{Z}_1 +\boldsymbol{A}_2\boldsymbol{Z}_2)
 +\boldsymbol{Z}_1^\top (\tau^{-1}\tau_h\boldsymbol{A}_1^\top \mm{A}_1 + \boldsymbol{R}(H_1) )\boldsymbol{Z}_1\\
    & &\quad \quad \quad \quad \quad +2\boldsymbol{Z}_1^\top (\tau^{-1}\tau_h\boldsymbol{A}_1^\top \boldsymbol{A}_2 )\boldsymbol{Z}_2
     +\boldsymbol{Z}_2^\top (\tau^{-1}\tau_h\boldsymbol{A}_2^\top \mm{A}_2 + \boldsymbol{R}(H_2) )\boldsymbol{Z}_2
    {\bigg)} {\bigg\}}
\end{eqnarray*}
The precision matrix of the stacked vector $\mm{u} = (\tilde{\mm{\epsilon}}^\top, \mm{Z}_1^\top,\mm{Z}_2^\top)^\top $ is then expressed by 
\begin{equation*}
    \boldsymbol{Q}_{\boldsymbol{u}} = \begin{pmatrix}
        \tau_h \boldsymbol{I} & -\tau^{-1/2}\tau_h\boldsymbol{A}_1 & -\tau^{-1/2}\tau_h \boldsymbol{A}_2 \\
        -\tau^{-1/2}\tau_h \boldsymbol{A}_1^\top & \tau^{-1}\tau_h\boldsymbol{A}_1^\top  \boldsymbol{A}_1 + \boldsymbol{R}(H_1) & \tau^{-1}\tau_h \boldsymbol{A}_1^\top \boldsymbol{A}_2\\
        -\tau^{-1/2}\tau_h \boldsymbol{A}_2^\top & \tau^{-1}\tau_h\boldsymbol{A}_2^\top \boldsymbol{A}_1 & \tau^{-1}\tau_h \boldsymbol{A}_2^\top \boldsymbol{A}_2 + \boldsymbol{R}(H_2)
    \end{pmatrix}.
\end{equation*}

\subsection{Modification for irregular time steps}\label{app:irregular}
To implement the proposed latent model component $\tilde{\mm{\epsilon}} = (\epsilon(t_1), \ldots , \epsilon(t_n))^\top$ for irregularly  spaced time points, we first define a high-resolution grid of $m>n$ equally spaced locations $\{s_j\}_{j=1}^m$ covering the range of time points $[\lfloor{t_1} \rfloor, \lceil{t_n}\rceil ]$. The  latent component at observation time  $t_i$ is then represented as 
$$\epsilon_i = \sum_{j=1}^m \mm{A}_{ij}s_j, \quad i=1,\ldots , n$$
where the matrix  $\mm{A}^{n\times m}$ contains interpolation weights based on the two grid points $s_j < t_i <s_{j+1}$, nearest to $t_i$. The non-zero elements of $\mm{A}$ are  $A_{ij} = s_{j+1}-t_i$ while $A_{i+1,j } = t_i - s_{j}$ for $ i=1,\ldots , n$ and $j = 1,\ldots , m-1$. This construction can be implemented in \texttt{R-INLA} by specifying the matrix $\mm{A}$ through the option \texttt{A.local}. It affects only the associated latent  model component, leaving other components of the linear predictor $\eta_i = E(y_i)$ unchanged. The mean component in \eqref{eq:regr-model} is modelled using the latent component \texttt{rw2} in \texttt{R-INLA}, which is readily defined for irregularly sampled using the \texttt{values} argument. 

\subsection{Additional results for the D-O events}\label{app:ar1-results}
\begin{figure}[ht]
\includegraphics[width = 0.99\textwidth]{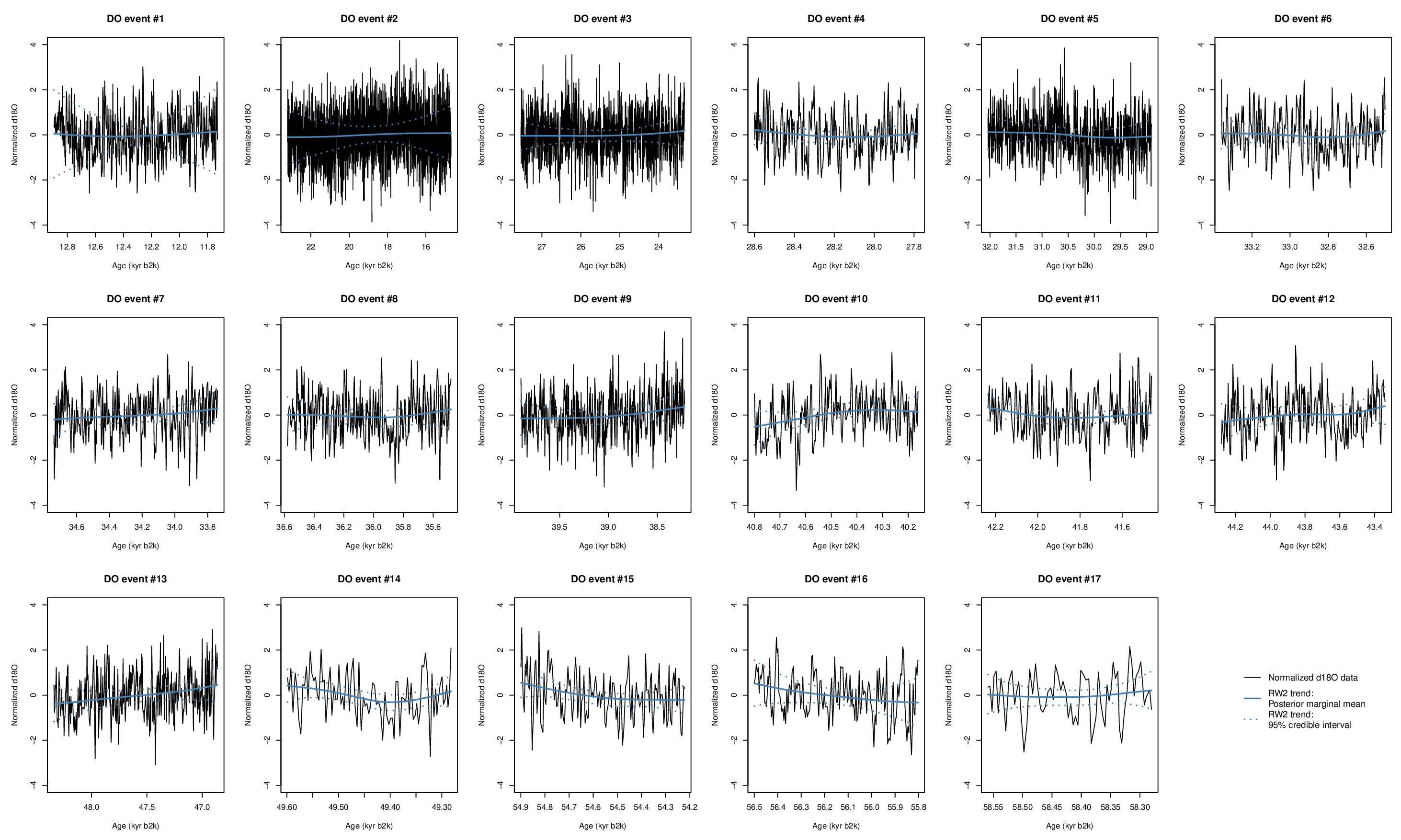}
\caption{D-O events: Standardized time series for the cold stadial phases, including the estimated posterior mean trend and $95\%$ credible intervals.} 
\label{fig:posterior-mean-ews}
\end{figure}

The posterior mean estimates of $\mu(.)$ in  \eqref{eq:regr-model} for each of the standardized time series are shown in Figure~\ref{fig:posterior-mean-ews}, displaying weak non-linear time-varying mean trends. Importantly, these are estimated jointly with the hyperparameters, avoiding a separate detrending step in which uncertainty of the mean trend would be ignored.  

A similar regression formulation was used in \citet{myrvoll1:25} and \citet{myrvoll2:25}, but modelling the noise term in \eqref{eq:regr-model} using time-varying AR(1) models. These models assume that the variance and lag-one autocorrelation depend on the common recovery rate $\lambda$ in  \eqref{eq:langevin}, thus assuming both increased variance and autocorrelation as $\lambda\rightarrow 0$. We have noticed in simulations that if this assumption is wrong, these time-varying AR(1) models are prone to falsely detecting increased autocorrelation in time series where changes are driven solely by increased marginal variance. To explore this further, we implemented a decoupled version of the model in \cite{myrvoll1:25}, modelling the marginal standard deviation separately using \eqref{eq:logitvariance}.

The resulting posterior probabilities of increasing lag-one autocorrelations using the different time-varying AR models
are summarized in Table~\ref{tab:do-events3}. The models in  \citet{myrvoll1:25} and \citet{myrvoll2:25} both flag a significant increase in the lag-one autocorrelation for
events 2, 5, 9, 11 and 16, having posterior probabilities above 0.95. Using the decoupled model, none of these are flagged
as significant while there is a high probability of increases in the marginal standard deviation, explaining the false alarms of increased lag-one autocorrelation. 

\begin{table}[ht]
\begin{center}
\begin{tabular}{rcccc}
\toprule
 no & AR(1) & Nested AR(1) & \multicolumn{2}{c}{Decoupled AR(1)} \\
  & $\hat p_{\phi}$ &$\hat p_{\phi}$ & $\hat p_{\phi}$ & $\hat p_{\sigma}$ \\
\midrule
1 & 0.883 & 0.891 &  0.419  & 0.861 \\
2 & \mm{0.970} & \mm{0.985} & 0.502 & \mm{0.972}  \\
3 & 0.520 & 0.335 &  0.860 & 0.335 \\
4 & $0.081$ & 0.030 &  0.247 & 0.203 \\
5 & $\mathbf{0.992}$ & \mm{0.999} &  0.197 & \mm{0.996} \\
6 & 0.245 & 0.178 & 0.458 & 0.300  \\
7 & 0.700 & 0.539 & 0.899 & 0.496  \\
8 & 0.895 & 0.890 & 0.893 & 0.709  \\
9 & \mm{0.956} & \mm{0.981} & 0.271 & \mm{0.947}  \\
10 & 0.107 & 0.082 & 0.000 & 0.240  \\
11 & \mm{0.957} & \mm{0.971} & 0.904 & 0.859  \\
12 & 0.030 & 0.019 & 0.688 & 0.078  \\
13 & 0.874 & 0.941 &  0.068 & 0.921 \\
14 & 0.745 & 0.752 & 0.542 & 0.629  \\
15 & 0.066 & 0.067 & 0.318 & 0.152 \\
16 & \mm{0.992} & \mm{0.996} & 0.289 & \mm{0.980}\\
17 & 0.606 & 0.595 & \mm{0.962} & 0.485 \\\hline
\end{tabular}
\caption{Posterior probability of increased autocorrelation $\hat p_{\phi}$ for the D-O events using the time-varying AR(1)-model in \citet{myrvoll1:25}, the nested AR(1) model in \citet{myrvoll2:25}, and a modified version of the model in \citet{myrvoll1:25} with decoupled variance and autocorrelation. For the decoupled model, we include the posterior probability of increased standard deviation $\hat p_{\sigma}$.} 
\label{tab:do-events3} 
\end{center}
\end{table}

\begin{figure}[ht]
\includegraphics[width = 0.99\textwidth]{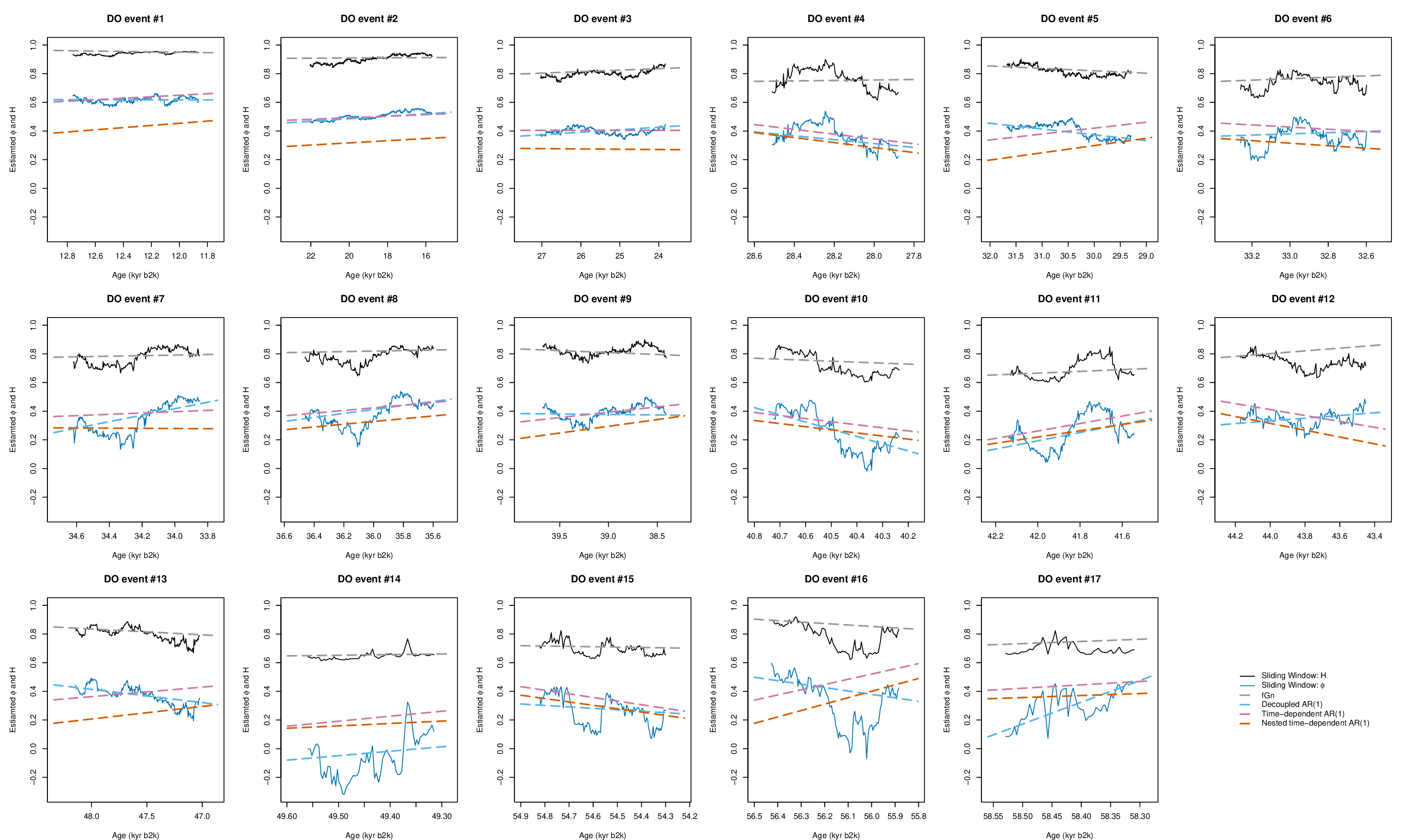}
\caption{D-O events: Running estimates for the Hurst exponents (solid black) and lag-one autocorrelations (solid blue), including the estimated change in the Hurst exponent using the proposed model (dashed grey) and the time-varying AR(1) models.
\label{fig:posterior-mean-H}}
\end{figure}

To further validate our results using the new time-varying fGn model for the D-O events, we compare the model-based estimates of $H$ with running estimates computed using sliding windows of length $n/4$. The running estimates are computed using the fGn-representation in \texttt{R-INLA}, accounting for irregularly sampled observations. As shown in Figure~\ref{fig:posterior-mean-H}, the increase in the Hurst exponent inferred by the proposed model is broadly consistent with the sliding window estimates, although the latter could be subject to substantial uncertainty and sensitivity to the choice of window. The agreement should therefore be interpreted as a qualitative consistency check, indicating that the model-based inference captures similar large-scale patterns as the sliding window estimates. 

Figure~\ref{fig:posterior-mean-H} also includes running estimates of the lag-one autocorrelation and the linear estimates fitting the models in  \citet{myrvoll1:25}, \citet{myrvoll2:25} and the decoupled version. The linear estimates using the modified decoupled model seem to match well with the sliding window estimates for all the series.  This is not the case using the original time-varying and nested AR(1) models, which in some cases give qualitatively different estimates than the other approaches. In conclusion, these result supports that we cannot claim to have early warning signals in terms of increased autocorrelation for these series. Also, an assumption of a common joint increase in both the variance and lag-one autocorrelation is not justified.  
\clearpage
\bibliographystyle{apalike}
\bibliography{shs}
\end{document}